\documentclass[conference]{IEEEtran}
\IEEEoverridecommandlockouts\IEEEpubid{\makebox[\columnwidth]{ 978-1-6654-8234-9/22/\$31.00 $\copyright$2022 IEEE \hfill}\hspace{\columnsep}\makebox[\columnwidth]{ }}

\usepackage{cite}
\usepackage{amssymb,amsfonts}
\usepackage{amsmath}
\makeatletter
\g@addto@macro\normalsize{%
  \setlength\abovedisplayskip{5pt}
  \setlength\belowdisplayskip{5pt}
  \setlength\abovedisplayshortskip{5pt}
  \setlength\belowdisplayshortskip{5pt}
}
\makeatother
\usepackage{algorithm, algorithmic}
\usepackage{graphicx}
\usepackage{textcomp}
\usepackage{xcolor}
\usepackage{bm}
\usepackage{amsthm}
\ifCLASSOPTIONcompsoc
  \usepackage[caption=false,font=normalsize,labelfont=sf,textfont=sf]{subfig}
\else
  \usepackage[caption=false,font=footnotesize]{subfig}
\fi
\usepackage{array}
\usepackage{bigstrut}
\usepackage{multirow}
\usepackage{booktabs}
\usepackage{threeparttable}
\usepackage{authblk}

\newtheoremstyle{exampstyle}
{0.35em plus 0.1em minus 0.1em }
  {0.35em plus 0.1em minus 0.1em}%
  {} 
  {} 
  {\bfseries} 
  {.} 
  {.5em} 
  {} 
\theoremstyle{definition}
\theoremstyle{exampstyle}\newtheorem{definition}{Definition}
\theoremstyle{exampstyle}\newtheorem{proposition}{Proposition}
\theoremstyle{exampstyle}\newtheorem{theorem}{Theorem}
\theoremstyle{exampstyle}\newtheorem{assumption}{Assumption}
\theoremstyle{exampstyle}\newtheorem*{remark}{Remark}
\theoremstyle{exampstyle}\newtheorem{lemma}{Lemma}

\newcommand{\rev}{\textcolor{black}}
\newcommand{\rrev}{\textcolor{black}}
\newcommand{\gai}{\textcolor{black}}
\newcommand{\ggai}{\textcolor{black}}

\def\BibTeX{{\rm B\kern-.05em{\sc i\kern-.025em b}\kern-.08em
    T\kern-.1667em\lower.7ex\hbox{E}\kern-.125emX}}
\begin{document}

\title{\rev{Centralized Network Utility Maximization with Accelerated Gradient Method\vspace{-1.5mm}
}\thanks{Corresponding authors are Zhiliang Wang and Xia Yin.}}
\author[*]{\rm Ying~Tian}
\author[$\dag$$\ddag$]{\rm Zhiliang~Wang}
\author[*$\ddag$]{\rm Xia~Yin}
\author[$\dag$]{\rm Xingang~Shi}
\author[$\dag$$\ddag$]{\rm Jiahai~Yang}
\author[$\dag$]{\rm Han~Zhang\vspace{-3.5mm}}
\affil[ ]{$^*$Department of Computer Science and Technology, BNRist, Tsinghua University, Beijing, China}
\affil[ ]{$^{\dag}$Institute for Network Sciences and Cyberspace, BNRist, Tsinghua University, Beijing, China}
\affil[ ]{$^{\ddag}$Zhongguancun Lab, Beijing, China}
\affil[ ]{E-mails:~y-tian18@mails.tsinguua.edu.cn, \{wzl,~shixg,~yang\}@cernet.edu.cn, \{yxia, zhhan\}@tsinghua.edu.cn\vspace{-3.5mm}
}

\maketitle

\begin{abstract}

Network utility maximization (NUM) is a well-studied problem for network traffic management and resource allocation.
Because of the inherent decentralization and complexity of networks, most researches develop decentralized NUM algorithms.
\rev{In recent years, the Software Defined Networking (SDN) architecture has been widely used, especially in cloud networks and inter-datacenter networks managed by large enterprises, promoting the design of centralized NUM algorithms.
To cope with the large and increasing number of flows in such SDN networks, existing researches about centralized NUM focus on the scalability of the algorithm with respect to the number of flows, however the efficiency is ignored.}
In this paper, we focus on the \rev{SDN} scenario, and derive a \rev{centralized, efficient and scalable} algorithm for the NUM problem.
By the designing of \rev{a} smooth utility function and \rev{a smooth} penalty function, we formulate the NUM problem with \rev{a} smooth objective function, which enables the use of Nesterov's accelerated gradient method.
We prove that the proposed method has $O(d/t^2)$ convergence rate, which is the fastest with respect to the number of iterations $t$, and our method is scalable with respect to the number of flows $d$ in the network.
Experiments show that our method obtains accurate solutions with less iterations, and achieves close-to-optimal network utility.

\end{abstract}


\section{Introduction}

As the number of network users and network traffic grow explosively, \rev{fair and efficient allocation of network resources becomes increasingly important. Regarding network resource allocation, a canonical and widely-studied problem} is the \emph{network utility maximization} (NUM) problem proposed by Kelly\cite{kelly1997}.
In \rev{the} NUM problem, there are sources sending elastic traffic along predetermined paths. Each source has a continuously differentiable, non-decreasing and concave utility function related to its sending rate, e.g.,  the most commonly used $\alpha$-fair function \ggai{proposed by Mo and Walrand\cite{ 10.1117/12.325891,879343}}. The objective is to maximize the sum of utility functions subject to link capacity constraints.
\rev{Because of the great generality and flexibility of the NUM problem, it has been combined with various network protocols and applied to different network scenarios like} wireless network\cite{1665005}, multi-layer radio access network\cite{9139398}, network function virtualization\cite{8673789} and so on.
Due to the inherent decentralization and complexity of networks, most of the NUM researches focus on developing decentralized algorithms\cite{kelly1998, 811451, 993307} based on dual decomposition\cite{1664999}. They assume that information like link capacity and flow route are only locally known by network devices\cite{10.1561/1300000007}, and the problem is  iteratively solved in a decentralized manner, where network devices locally solve their subproblems and communicate about necessary information like intermediate results. \ggai{Decentralized methods} usually scale with the number of concurrent flows $d$, and the fastest convergence rate achieved by existing decentralized methods is $O(1/t)$ \cite{16M1059011, 6756941} \ggai{where $t$ is the number of iterations}.

In recent years, Software Defined Networking (SDN)\cite{Foundation2012Software} draws great attention because of its ability to decouple control plane from data plane.
\rrev{Unlike traditional decentralized networks,  in the SDN architecture there is an SDN controller has a global view about network topology and network states,} \rrev{thus} it \rrev{can} centrally computes forwarding paths, flow rates, etc., and \rrev{then} controls the behavior of switches in the data plane.
The SDN architecture \rev{has been widely used, especially in cloud networks and inter-datacenter networks managed by large enterprises\cite{Jain2013B4, Hong2013Achieving}, where a large number of flows are competing for scarce and expensive network resources like link bandwidth.
To serve the urgent need of managing and optimizing \ggai{resource} allocation in such networks, it is essential to develop \textit{centralized}, \textit{efficient} and \textit{scalable} algorithms for the NUM problem:  the algorithm should be
\textit{centralized} to \rrev{solve the problem} on the SDN controller \rrev{with global information}, be \textit{efficient} to achieve near-optimal results fast or even in real time, and be \textit{scalable} to accommodate the large and even increasing number of network flows.}

Existing researches \rrev{about centralized NUM algorithm} mainly focus on the scalability of the algorithm\rev{,  but the efficiency is not ideal.}
Gupta et al.\cite{7492926} \rev{achieve scalability by converting the original} NUM problem into a simpler problem involving only aggregate flows, \rev{without guaranteeing the convergence rate and efficiency of the algorithm.}
The state-of-the-art work Exp-NUM\cite{8737600} \rev{solves the penalty-based NUM formulation with exponentiated gradient descent. The algorithm is proven to be scalable with respect to the number of flows. But it shows poor efficiency because it converges quite slow with respect to the number of iterations $t$, and needs a lot of iterations to obtain an accurate solution.}

In this paper, we follow the classical formulation of the NUM problem with $d$ concurrent flows and \rev{the $\alpha$-fair utility function}, and derive a centralized \rev{\textit{accelerated gradient method} (AGM)} for the NUM problem, \rev{which is both efficient and scalable.
While most researches only exploit the Lipschitz continuous property of the $\alpha$-fair function, we go a step further by revealing the smoothness of the $\alpha$-fair function} and design an appropriate smooth penalty function for the link capacity constraints, enabling the use of Nesterov's accelerated gradient method\cite{nesterov1983method}, which is believed to be the fastest iterative algorithm for smooth objective function. \rev{Our method is proven to have the fastest convergence rate with respect to the number of algorithm iterations $t$ which shows great efficiency, and is scalable with respect to the number of flows $d$, showing its great potential for resource allocation in large-scale centralized controlled networks.}

\begin{table*}
\centering
\renewcommand\arraystretch{1.1}
\caption{Network type, convergence rate, iteration complexity, and objective function class of the first-order methods\vspace{-1.5mm}}
    \begin{tabular}{|p{1.8cm}<{\centering}|c|c|c|p{5.4cm}<{\centering}|}\hline
\textbf{Network Type}   & \textbf{Method}& \textbf{Convergence Rate}  & \textbf{Iteration Complexity} &{\textbf{Objective Function Class}$^{\rm a}$}\\ \hline
    \multirow{3}{1.8cm}{\centering Decentralized controlled }   &  Dual gradient\cite{NESTEROV2018907} &  $O(1/\sqrt{t})$  & $\Omega(1/\varepsilon^2)$ &{concave} \\ \cline{2-5}
        &   Virtual queue\cite{16M1059011} &  $O(1/t)$  & $\Omega(1/\varepsilon)$ & {concave (\rev{and} Lipschitz continuous constraint)} \\ \cline{2-5}
            & Primal-dual\cite{6756941} & $O(1/t)$  & $\Omega(1/\varepsilon)$ & {strongly concave} \\ \hline
     \multirow{4}{1.8cm}{\centering SDN network}  & \multirow{2}{*}{{Projected gradient ascent{$^{\rm b}$}}\cite{MAL-050}} & $O(\sqrt{d}/\sqrt{t})$  & $\Omega(d/\varepsilon^2)$ &{concave, Lipschitz continuous} \\  \cline{3-5}
       && $O(d/t)$  & $\Omega(d/\varepsilon)$ & {concave, smooth} \\  \cline{2-5}
        & Exp-NUM\cite{8737600} & $O(\sqrt{\ln{d}}/\sqrt{t})$  & $\Omega(\ln{d}/\varepsilon^2)$ &{concave, Lipschitz continuous} \\  \cline{2-5}
                 & \textit{AGM (this paper)} & $O(d/t^2)$  & $\Omega(\sqrt{d}/\sqrt{\varepsilon})$ &\textit{concave, smooth} \\ \hline
        \end{tabular}
        \begin{tablenotes}
          \footnotesize
          \item $^{\rm a }$ In this table, we assume that the NUM problem is a maximizing problem so all objective functions should be concave.
           \item $^{\rm b }$ Projected gradient ascent has different convergence rate and iteration complexity for Lipschitz continuous and smooth objective functions.\vspace{-4mm}
        \end{tablenotes}
    \label{relatedw}
    \end{table*}

The main contributions of our paper are as follows:
 \begin{itemize}
 \item We are the first to formulate the NUM problem with a smooth objective function. First, we propose and prove \rev{that the $\alpha$-fair utility function becomes smooth with only minor changes.}
 Then, we design a smooth penalty function for the penalty-based formulation, which leads to the overall smoothness of the objective function in cooperation with the smooth $\alpha$-fair utility function or any other smooth utility functions. The NUM formulation with smooth objective function benefits the use of existing algorithms\rev{,} and will encourage the design of other fast algorithms.

 \item We \rev{propose} to apply Nesterov's accelerated gradient method (AGM) to solve \rev{the} centralized NUM problem. Thanks to the smoothness of the objective function, the algorithm has $O(d/t^2)$ convergence rate \ggai{under the assumption that the number of network links is much smaller than $d$}, which is the fastest with respect to $t$ \gai{compared with all existing decentralized and centralized methods}. The iteration complexity of our algorithm is proportional to $\sqrt{d}$, so the algorithm \rev{is scalable with respect to the number of flows} and can maintain good performance as $d$ increases. Besides, AGM is guaranteed to achieve more accurate solutions than \rev{the state-of-the-art method} Exp-NUM with enough number of iterations. We also use two restart heuristics to boost the convergence of AGM and alleviate the bumps during the iteration.

\item We conduct thorough experiments to evaluate the algorithms and \rev{demonstrate the efficiency and scalability of AGM.} The experimental results are in good agreement with the theoretical results. AGM has the fastest convergence rate and achieves accurate solutions. Besides, our method obtains close-to-optimal utility function value, indicating the effectiveness of the smoothly penalized NUM formulation. And our method performs consistently well with different number of flows, showing its good scalability with respect to $d$.
 \end{itemize}

The rest of the paper is structured as follows. Section~\ref{relatedwork} is the related work. Section~\ref{formulation} presents the formulation of the NUM problem. In Section~\ref{gdsec}, we introduce the state-of-the-art gradient based methods for centralized NUM problem.
We formulated the NUM problem with smooth objective function in Sec~\ref{smoothform}, and then present the accelerated gradient method in Sec~\ref{agmsec}.
In section~\ref{evaluation}, we evaluate the proposed algorithm. \gai{In section~\ref{discussion}, we discuss the limitations of our method.}
Finally, we make a conclusion in Section~\ref{conclusion}.


\section{Related Work}\label{relatedwork}

In the past two decades, there have been \ggai{plenty} of research in NUM. \ggai{Most of them study concave utility functions, and some specifically use $\alpha$-fair function in the problem formulation or as a representative example\cite{kelly1998,811451, 16M1059011,6756941,9139398, 8064335,7492926,8737600}.} NUM algorithms are usually iterative, whose performance is related to: i) \emph{convergence rate}, i.e., the achieved error $\varepsilon$\footnote{The error $\varepsilon$ is measured by the gap between the objective value of the current solution and the optimal value.} at step $t$, or equivalently, \emph{iteration complexity}, i.e., the number of iterations required to achieve an error; ii) \emph{per-iteration cost}, measured by the computational complexity or number of mathematical operations per iteration.

Newton's method is believed to have the fastest convergence rate. Wei et al.\cite{5718026} proposed a distributed Newton's method for \rev{the} NUM problem. Zargham et al.\cite{6831412, 6676846} accelerated dual decent with an approximate Newton-like algorithm. However, the convergence rate is not concisely derived. Besides, Newton's method requires the computation of Hessian matrix and its inverse, causing heavy per-iteration cost.

\rev{Therefore}, first-order methods that only require the computation of the first derivative (gradient) are preferred. Kelly et al.\cite{kelly1998} decomposed the NUM problem into a user subproblem and a network subproblem and solve the problem with gradient descent. Low et al. \cite{811451} proposed another decomposition framework and consider asynchronous situation. Both works leverage the dual-based gradient descent algorithm, but concise proof of the convergence rate is not given. In \cite{NESTEROV2018907}, authors proved that the dual gradient method with primal averaging has $O(1/\sqrt{t})$  convergence rate. In \cite{16M1059011}, authors assumed that constraint functions are Lipschitz continuous and convex, and proposed a virtual queue algorithm based on dual gradient method, achieving $O(1/t)$ convergence rate. Assuming strong concavity of the utility function, Beck et al.\cite{6756941} solved the dual problem with accelerated gradient method. But their primal-dual method only achieves $O(1/t)$ convergence rate for the primal solution.

The works mentioned above all develop distributed algorithms for decentralized controlled networks through dual decomposition. As the network architecture evolves, NUM algorithms considering other network architectures emerge. Karako\c{c} et al.\cite{9139398} proposed a multi-layer decomposition for NUM problem \rev{in the radio access network solved by dual gradient descent method.} Allybokus et al. \cite{8064335} considered an SDN network with multiple distributed controllers, and developed a distributed algorithm based on alternating direction method of multipliers (ADMM) algorithm.

For \rev{the} NUM problem in centralized SDN networks, a classic method is projected gradient ascent\cite{MAL-050} (or or projected gradient descent for minimizing problem\rev{s}). For Lipschitz objective functions, it has $O(\sqrt{d}/\sqrt{t})$ convergence rate and $\Omega(d/\varepsilon^2)$ iteration complexity. For smooth objective functions, its performance guarantee is improved to $O(d/t)$ convergence rate and $\Omega(d/\varepsilon)$ iteration complexity. Because the iteration complexity of projected gradient ascent is proportional to $d$, it \rev{is not scalable with respect to} the number of flows. To cope with \rev{the increasing} number of flows, Gupta et al. \cite{7492926} proposed a NUM decomposition in which the set of flows are grouped by source-destination pairs, and solved the problem with ADMM algorithm. But the convergence rate is not studied. Exp-NUM\cite{8737600} is most relevant to our paper. Exp-NUM uses a penalty-based formulation of the NUM problem and additionally adds a global flow rate bound, which allows the problem to be solved with exponentiated gradient descent, i.e., mirror descent\cite{nemirovskij1983problem}. The advantage of  Exp-NUM is that the iteration complexity is $\Omega(\ln d/\varepsilon^2)$, thus the algorithm is scalable with respect to the number of flows $d$. However, the convergence rate of Exp-NUM is $O(\sqrt{\ln{d}}/\sqrt{t})$, implying \rev{that it cannot converge to accurate solutions efficiently.}

The convergence rate of our method AGM is $O(d/t^2)$, which is the fastest w.r.t. $t$ compared with all \gai{decentralized and centralized} methods mentioned above \footnote{\gai{We believe AGM converges faster than Gupta et al. \cite{7492926} in terms of $t$ because $O(1/t^2)$ is recognized as the fastest convergence rate of all known first-order methods.}}. Besides, the iteration complexity is $\Omega(\sqrt{d}/\sqrt{\varepsilon})$ in the worst case, i.e., the number of iterations required for an $\varepsilon$-optimal solution is proportional to the square root of the number of flows. \rev{Thus, our method is both efficient and scalable.}
In Table~\ref{relatedw}, we summarized the network type, convergence rate, iteration complexity and objective function class of the first-order methods.


\vspace{-0.5mm}

\section{Network Utility Maximization Problem}\label{formulation}

\vspace{-1mm}

In this section, we present the formulation of the NUM problem and introduce the commonly used utility function and penalty function. The notations used in this paper is summarized in Table~\ref{notation}.

\vspace{-0.5mm}
\subsection{Problem Formulation}
\vspace{-1mm}

The system model of the NUM problem consists of the following aspects:

\textit{1) Network:} The network is represented by a graph $G=(V, E)$, where $V$ is the router set and $E$ is the unidirectional link set. Each link $e \in E$ has a capacity $c_e$.

\textit{2) Flows:}
Let $S$ represents the set of elastic flows whose sending rate can be controlled by the source or controller. $s \in S$ represents a specific flow, and $x_s$ is the rate of $s$. We also use $\bm{x}=(x_s)_{s\in{S}}$ as the column vector of flow rates. Besides, let $d$ be the number of flows in the network, i.e., $d=|S|$, $\bm{x}\in\mathbb{R}^d_+$.

\textit{3) Routing path:}
The routing path is predetermined for each flow $s \in S$. We assume that there is one path for each flow and the $|E| \times |S|$ routing matrix $\mathbf{A}$ is defined as:
\vspace{-1mm}
\begin{equation}\label{A}
\mathbf{A}_{e, s}=\left\{
\begin{array}{rcl}
1, && {\text{if flow $s$ passes through link $e$}}\\
0, && {\text{otherwise}}\\
\end{array} \right.
\end{equation}
This definition is suitable for simple paths in which a flow $s$ fully passes each link. For paths in which a flow may split over links, e.g., Equal-cost multi-path (ECMP) or Segment Routing\cite{7417124} paths, $\mathbf{A}_{e,s}$ can be defined as the fraction of flow $s$ passes through link $l$ and relaxed to $[0,1]$. For convenience, we use the row vector $\mathbf{A}_e$ to denote the $e$-th row of matrix $\mathbf{A}$.

\textit{4) Utility function:}
Each flow $s$ is associated with a \textit{flow utility function} $U_s:\mathbb{R}_+\rightarrow\mathbb{R}$, i.e., each flow $s$ gains a utility $U_s(x_s)$ when it sends data at rate $x_s$. We call $U(\bm{x}) = \sum_{s\in S}U_s(x_s)$ the \textit{network utility function}.

The NUM problem is formulated as:

\vspace{-2mm}
\begin{equation}
  \label{num}\tag{NUM}
  \begin{aligned}
    \max \,\,\,&U(\bm{x})={\sum_{s \in S}U_s(x_s)}\\
    \vspace{-2mm}
       \text{s.t.,}\,\,\,\, &\mathbf{A}_e \bm{x} \le c_e ,\,\, \forall e \in E\\
     &x_s \ge 0,\,\,\forall s\in S
  \end{aligned}
\end{equation}
\vspace{-3mm}

\begin{table}
\vspace{-1mm}
\centering
\renewcommand\arraystretch{1}
\caption{\vspace{-0.5mm}Notations\vspace{-1.5mm}}
    \begin{tabular}{|c |l|}\hline
    \textbf{Symbol} & \multicolumn{1}{c|}{\textbf{Description}}\\ \hline
      $G=(V,E)$ & \multicolumn{1}{m{5.1cm}|}{The network graph, where $V$ is the node set and $E$ is the unidirectional link set. } \\ \hline
           $c_e$ & \multicolumn{1}{m{5.1cm}|}{The capacity of link $e$. }\\  \hline
    $S$ &\multicolumn{1}{m{5.1cm}|}{ The set of traffic flows. }\\\hline
        $d$ &\multicolumn{1}{m{5.1cm}|}{The number of flows, i.e.,  $d = |S|$. }\\\hline
    $\bm{x}$ &\multicolumn{1}{m{5.1cm}|}{The vector of flow rates where $x_s$ is the rate of flow $s$.}\\ \hline
        $\bm{x}^{(t)}$   & \multicolumn{1}{m{5.1cm}|}{The obtained solution at iteration $t$.} \\\hline
        $\bm{x}^{*}$   & \multicolumn{1}{m{5.1cm}|}{The optimal solution.} \\\hline
    $\mathbf{A}$  & \multicolumn{1}{m{5.1cm}|}{The routing matrix where $\mathbf{A}_{e,s}$ represents whether flow $s$ passes link $e$ or not.}\\ \hline
     $U_s, U$  & \multicolumn{1}{m{5.1cm}|}{The flow utility function and the network utility function, where $U=\sum_{s\in S}U_s$.}\\ \hline
    $P_e, P$ &\multicolumn{1}{m{5.1cm}|} {The link penalty function and the network penalty function, where $P = \sum_{e\in E} P_e$.} \\\hline
    $V$ & \multicolumn{1}{m{5.1cm}|}{The objective function.} \\\hline
     $V^*$ & \multicolumn{1}{m{5.1cm}|}{The optimal objective function value.} \\\hline
         $\mathbb{R}^d_+$ & \multicolumn{1}{m{5.1cm}|}{The space of $d$-dimensional nonnegative column vector.} \\\hline
         $[\,\,\cdot\,\,]_+$ & \multicolumn{1}{m{5.1cm}|}{The  projection on $\mathbb{R}^d_+$.} \\\hline
         $\Vert \cdot \Vert$ & \multicolumn{1}{m{5.1cm}|}{The Euclidean norm.} \\\hline
         $\odot$  & \multicolumn{1}{m{5.1cm}|}{The Hadamard product.} \\\hline
    \end{tabular}
    \label{notation}
    \end{table}

The objective is to maximize the network utility function. It is constrained that the traffic carried by a link should not exceed its capacity, and the flow rate should be nonnegative. Usually, the utility function is defined by service provider in order to achieve a certain type of resource allocation. The $\alpha$-fair function\cite{879343} is a well-known and most commonly used flow utility function, defined as Eq.~\eqref{util}. The $\alpha$-fair function is flexible to capture different optimization goals, and different values of \rev{the parameter} $\alpha$ yields different fairness criteria. For example, $\alpha=0$ leads to the classic throughput maximization, $\alpha=1$ yields proportional fairness\cite{kelly1997}, $\alpha=2$ yields to potential delay fairness\cite{752159} and $\alpha \to \infty$ gives min-max fairness.
\vspace{-1mm}
\begin{equation}
U_{\alpha}(x_s)=\left\{
\begin{array}{rcl}
&\!\!\!\!\!\!\!\!\dfrac{x_s^{1-\alpha}}{1-\alpha},      & {\alpha \ge0, \alpha \neq 1}\\
&\!\!\!\!\!\!\!\!\ln{x_s},   & {\alpha=1}\\
\end{array} \right.\label{util}
\end{equation}

\ggai{For the sake of our proposed method AGM and its associated theorems and proofs, we make the following assumptions:
\begin{assumption}\label{ass_E}
The network topology is sparse enough so that $|E|\ll|V|(|V|-1)$. This assumption holds when network topologies are not full-mesh and are much sparser (e.g., the topologies in Topology Zoo dataset\cite{6027859}, or G\'EANT and RF1221 used in \ref{evaluation} Evaluation). 
\end{assumption}
\begin{assumption}\label{ass_D}
The number of flows $d$ is at least the same order of magnitude as the number of node pairs $|V|(|V|-1)$. $d$ could be greater than $|V|(|V|-1)$ because we consider fine-grained flows sent by network users rather than aggregate flows between source-destination pairs.
\end{assumption}
\begin{assumption}\label{ass_U}
The utility function is concave and smooth. The $\alpha$-fair function is well-known to be concave, and we will reveal its smoothness later in the paper.
\end{assumption}}

\subsection{Penalty-based Formulation}

In problem \eqref{num}, the link capacity constraints form a polyhedron and is hard to handle. For gradient based methods, projection onto the polyhedron can be time consuming. One way to deal with it is to derive the Lagrange dual of \eqref{num}, which is suitable for decentralized algorithms.
Another conventional way is to relax the problem by using a penalty-based formulation\cite{10.1561/1300000007}, and the objective function becomes:

\vspace{-2mm}
\begin{equation}
V(\bm{x})={\sum_{s \in S}U_s(x_s)}-\mu\sum_{e\in E}P_e(\mathbf{A}_e \bm{x})\nonumber
\end{equation}
\vspace{-2.5mm}

The function $P_e: \mathbb{R}_+ \to \mathbb{R}$ is a \textit{link penalty function} that penalizes the arriving rate for overshooting the link capacity. We call $P(\bm{x}) = \sum_{e\in E}P_e(\mathbf{A}_e \bm{x})$ the \textit{network penalty function}. $\mu>0$ is a penalty parameter that balances the utility function and penalty function. To accommodate the convention of using gradient descent (instead of ascent) based algorithms for minimization (instead of maximization) problems, we negate the function $V(x)$, and then the problem is formulated as:
\vspace{-0.5mm}
\begin{equation}\label{nump}\tag{NUM-P}
  \begin{aligned}
    \min \,\,\,V_P(\bm{x})=&-{\sum_{s \in S}U_s(x_s)}+\mu\sum_{e\in E}P_e(\mathbf{A}_e \bm{x})\\
    \vspace{-1mm}
  \text{s.t.,}\,\,\,\, & x_s \ge 0,\,\, \forall s\in S
  \end{aligned}\!\!\!\!\!\!\!\!\!\!\!\!\!\!\!\!
      \vspace{-0mm}
\end{equation}

The optimization objective now is to minimize the negative of the network utility functions plus the network penalty function. The advantage of the problem \eqref{nump} is that the link capacity constraints are removed so that the feasible region now becomes the nonnegative orthant $\mathbb{R}^d_+$ instead of a polyhedron, making the problem easier to deal with. Besides, by using an appropriate penalty function and parameter $\mu$, the exact value of the optimal solution of \eqref{num} can be obtained through solving \eqref{nump}\cite{10.1561/1300000007, 8737600}.

 A typical choice for the link penalty function is to use the exact penalty function \cite{10.2307/2627851} $P_{e, +}(\mathbf{A}_e \bm{x})$  as  shown in Eq.~\eqref{exactp}. The penalty value is equal to how much the link capacity is exceeded. However, this function is not continuously differentiable when $\mathbf{A}_e \bm{x}=c_e$ , which may cause numerical issues and make the problem hard to handle.
 \vspace{-1mm}
 \begin{equation}\label{exactp}
P_{e,+}(\mathbf{A}_e \bm{x}) = \max \{\mathbf{A}_e \bm{x}-c_e,0\}
\end{equation}


\section{\vspace{1mm}Gradient Methods for NUM with The Exact Penalty Function}\label{gdsec}

\begin{algorithm}[tb]
\renewcommand{\algorithmicrequire}{\textbf{Input:}}
\renewcommand{\algorithmicensure}{\textbf{Output:}}
\caption{Projected Gradient Descent (PGD)}\label{pgd}
\begin{algorithmic}[1]
\REQUIRE initial point $\bm{x}^{(0)}$, step size $\gamma$, number of iterations $T$
\ENSURE final solution $\hat{\bm{x}}^{(T)}$
\FOR{$t \leftarrow 0$ \textbf{to} ${T-1}$}
    \STATE $\bm{x}^{(t+1)} \leftarrow \left[\bm{x}^{(t)} - \gamma  \nabla V_P(\bm{x}^{(t)})\right]_+$
\ENDFOR
\RETURN  $\hat{\bm{x}}^{(T)} \leftarrow \frac{1}{T}\sum_{t=1}^{T}\bm{x}^{(t)}$
\end{algorithmic}
\end{algorithm}

In this section, we introduce the classic projected gradient descent (PGD)\cite{MAL-050} and the state-of-the-art method Exp-NUM\cite{8737600} for NUM problem with the $\alpha$-fair utility function $U_{\alpha}$ \rev{(Eq.~\eqref{util})} and exact penalty function $P_{e,+}$ \rev{(Eq.~\eqref{exactp})}. We describe the prerequisites, algorithm procedures,  performance guarantees of these methods, and analyze their drawbacks.

The pseudocode of PGD is shown in Algorithm~\ref{pgd}. At each iteration, PGD takes a step in the opposite direction of the gradient, and then projects the obtained point back onto $\mathbb{R}^d_+$. The final solution $\hat{\bm{x}}^{(T)}$ is the average over all intermediate points in the iteration. To give the convergence rate of PGD, an assumption is made:

\begin{assumption}\label{ass_L}
The partial derivatives of $V_P(\bm{x})$ are bounded, i.e., there exists $M\ge0$ so that
 \begin{equation}
 \left|\frac{\partial V_p(\bm{x})}{\partial x_s}\right|\le M,  \,\,\,\,\, \forall \bm{x}\in \mathbb{R}_+^d, s \in S,\nonumber
 \end{equation}
 implying that $V_p(\bm{x})$ is $(M\sqrt{d})$-Lipschitz continuous\footnote{See Definition~\ref{lipschitz} and Lemma~\ref{boundedgradient} in Section~\ref{preliminaries}.}.
\end{assumption}

Then PGD achieves the following convergence guarantee:

\begin{proposition}[Convergence rate of PGD\cite{MAL-050}] \label{prop_gd}
If Algorithm~\ref{pgd} is run for $T$ iterations with initial point $\bm{x}^{(0)}$, step size $\gamma = \frac{\Vert \bm{x}^{*}\!-\bm{x}^{(0)} \Vert }{ M\sqrt{dt}}$, then the solution  $\hat{\bm{x}}^{(T)}$  satisfies:
\begin{equation}
V_P(\hat{\bm{x}}^{(T)}) - V_P^{*} \le \frac{M\sqrt{d}\,\Vert \bm{x}^{*}\!\!-\bm{x}^{(0)} \Vert }{\sqrt{T}},\nonumber
\end{equation}
i.e., the error is $\varepsilon= O(\sqrt{d}/\sqrt{t})$ at step $t$, or equivalently, PGD achieves an $\varepsilon$-optimal solution with $\Omega(d/\varepsilon^2)$ iterations.
\end{proposition}

According to Proposition~\ref{prop_gd}, \rev{PGD is neither efficient nor scalable:} numerous iterations are required to reach an accurate solution \rev{because of  the slow convergence rate with respect to $t$}, and \rev{the iterative complexity grows proportionally to the number of network flows $d$.}


Exp-NUM\cite{8737600} is the state-of-the-art gradient method for the NUM problem. Considering problem \eqref{nump} with \rev{the $\alpha$-fair utility function $U_{\alpha}$ and exact penalty function $P_{e,+}$}, authors of Exp-NUM propose an exponentiated gradient descent method whose iteration complexity \rev{is nearly independent of} the number of flows. \rev{In Exp-NUM, besides} Assumption~\ref{ass_L}, another assumption is made:

\begin{assumption}\label{ass_C}
The sum of flow rates is bounded by a parameter $C$, thus the feasible region of \eqref{nump} becomes $\mathcal{X} = \{ \bm{x} \in{\mathbb{R}^d_+}:   x_s\ge 0~\text{and}~\sum_{s\in S}x_s\le C\}$,  where  $C \ge 0$.
\end{assumption}

\begin{algorithm}[tb]
\renewcommand{\algorithmicrequire}{\textbf{Input:}}
\renewcommand{\algorithmicensure}{\textbf{Output:}}
\caption{Exponentiated Gradient Descent (Exp-NUM)}\label{expnum}
\begin{algorithmic}[1]
\REQUIRE  initial point $\bm{x}^{(0)}$, step size $\gamma$, number of iterations $T$,  global flow rate bound $C$
\ENSURE final solution $\hat{\bm{x}}^{(T)}$
\FOR{$t \leftarrow 0$ \textbf{to} ${T-1}$}
   \STATE $\bm{v}^{(t)} \leftarrow \nabla V_P(\bm{x}^{(t)})$
    \STATE $\bm{x}^{(t+1)} \leftarrow C\dfrac{\bm{x}^{(t)} \odot \exp(-\gamma \bm{v}^{(t)})}{C+\sum_{s\in S}x^{(t)}_s[\exp(-\gamma \bm{v}^{(t)})-1]}$
\ENDFOR
\RETURN  $\hat{\bm{x}}^{(T)} \leftarrow \frac{1}{T}\sum_{t=1}^{T}\bm{x}^{(t)}$
\end{algorithmic}
\end{algorithm}

The pseudocode of Exp-NUM  are shown in Algorithm~\ref{expnum}. At each iteration, it updates $\bm{x}^{(t+1)}$ with an exponential function of the gradient. The final solution is still the averaged point $\hat{\bm{x}}^{(T)}$. Exp-NUM has the following convergence rate guarantee:

\begin{proposition}[Convergence rate of Exp-NUM\cite{8737600}] \label{prop_expnum}
If Algorithm~\ref{expnum} is run for $T$ iterations with initial point $\bm{x}^{(0)}$ and step size $\gamma = M^{-1}\sqrt{2D(\bm{x}^{*}, \bm{x}^{(0)})/t}$, then the solution  $\hat{\bm{x}}^{(T)}$  satisfies:
\begin{equation}
V_P(\hat{\bm{x}}^{(T)}) - V_P^{*} \le \frac{M\sqrt{2CD(\bm{x}^{*}, \bm{x}^{(0)})}}{\sqrt{T}},\label{expnumconvergeeq}
\end{equation}
where $D(\bm{x}^{*}, \bm{x}^{(0)})$ is the Bregman divergence\cite{8737600} between $\bm{x}^{*}$ and $\bm{x}^{(0)}$. If $\bm{x}^{(0)}$ is the barycenter of $\mathcal{X}$, then $D(\bm{x}^{*}, \bm{x}^{(0)})$ is less than $C\ln{(d+1)}$, and ExpNUM achieves $\varepsilon= O(\sqrt{\ln{d}}/\sqrt{t})$ error at step $t$, or equivalently,  Exp-NUM achieves an $\varepsilon$-optimal solution with $\Omega(\ln{d}/\varepsilon^2)$ iterations.
\end{proposition}


There are a few things to note about PGD and Exp-NUM:

1) \rev{Both PGD and Exp-NUM are not efficient and need a lot of iterations to converge.} The advantage of Exp-NUM compared with PGD is that the iteration complexity  is in logarithmic time relative to $d$, thus the algorithm is scalable with respect to the number of flows. However, its convergence rate is still proportional to $1/\sqrt{t}$.

2) PGD and Exp-NUM uses the average point $\hat{\bm{x}}^{(T)}$ as the final solution, instead of the point at the last iteration $\bm{x}^{(T)}$. The reason is that the intermediate solutions at each iteration may oscillate around the optimal solution and every-step progress cannot be guaranteed, while using the average point yields a provable convergence theorem.

3) \ggai{Assumption~\ref{ass_L} and Assumption~\ref{ass_C}} are not valid, \rev{so the convergence analyses of PGD and Exp-NUM, i.e., Proposition~\ref {prop_gd} and  Proposition~\ref{prop_expnum}, are not solid.}
 Assumption~\ref{ass_L} does not hold for $V_P(\bm{x})$ with the $\alpha$-fair utility function $U_{\alpha}$ and exact penalty function $P_{e,+}$.
 First, $P_{e,+}(\mathbf{A}_e\bm{x})$ is non-differential when $\mathbf{A}_e \bm{x}=c_e$, so $\frac{\partial V_p(\bm{x})}{\partial x_s}$ cannot be calculated.
 Second, even if it is artificially specified that $\frac{\partial P_{e,+}(\mathbf{A}_e\bm{x})}{\partial x_s}=0$ when $\mathbf{A}_e \bm{x}=c_e$, still there is no upper bound $M$ on  $\left|\frac{\partial V_p(\bm{x})}{\partial x_s}\right|$.
By Eq.~\eqref{util}, ${U_s'(x_s)}= x_s^{-\alpha}$. So when $x_s = 0$, ${U_s'(x_s)} = \infty$, and $\left|\frac{\partial V_p(\bm{x})}{\partial x_s}\right| = \infty$. Thus the partial derivatives of $V_P(\bm{x})$ over $\bm{x} \in \mathbb{R}_+^d$ (or $\bm{x} \in \mathcal{X}$) are not bounded.
Exp-NUM chooses to add a small number to the argument $x_s$ of  $U_{\alpha}(x_s)$ to handle this problem in the experiments, but no theoretical analysis is given. (See the next section for detail.)
Assumption~\ref{ass_C} alters the feasible region of \eqref{nump}. The optimal solution of \eqref{nump} is not changed only if $C\ge\sum_{e\in E}c_e$.



\section{NUM Problem with Smooth Objective Function} \label{smoothform}

In this section, we propose the NUM problem formulation with smooth objective function. We start with some preliminary definitions and lemmas. And then we design a smooth utility function and a smooth penalty function respectively to construct the smooth objective function. The obtained NUM problem benefits the use of PGD and Exp-NUM algorithms. More importantly, it enables the use of \rev{the accelerated gradient method} presented in the next section.

\subsection{Preliminaries}\label{preliminaries}

We define the concept \textit{Lipschitz continuous} and \textit{smooth} with respect to the Euclidean norm $\Vert\cdot\Vert$, and introduce some relevant properties. The proof of the lemmas is trivial and straightforward, so it is omitted.

\begin{definition}[Lipschitz continuous]\label{lipschitz}
A function $f$ is $L$-\textit{Lipschitz continuous} (or $L$-\textit{Lipschitz} for short) on $Q\in \mathbb{R}^n$ with \textit{Lipschitz constant} $L \ge 0$, if for all $\bm{x}, \bm{y} \in Q$, we have
\begin{equation}
\Vert f(\bm{x})-f(\bm{y}) \Vert \le L \Vert \bm{x} - \bm{y} \Vert\nonumber
\end{equation}
\end{definition}

\begin{lemma} \label{boundedgradient}
A function $f$ is $L$-Lipschitz on $Q\in \mathbb{R}^n$  if and only if there exists $L\ge0$ so that $\Vert \nabla f(\bm{x})\Vert \le L$ for all  $\bm{x} \in Q$.
\end{lemma}

\begin{lemma} \label{compositionlipschitz}
The composition of Lipschitz functions is Lipschitz. Specifically, if a function $f:X\rightarrow Y$ is $L_f$-Lipschitz on $X$, and a function $g: Y\rightarrow Z$ is $L_g$-Lipschitz on $Y$, then the function $g\circ f:X\rightarrow Z$ is $(L_fL_g)$-Lipschitz on $X$.
\end{lemma}

\begin{definition}[Smooth]\label{smooth}
A function $f$ is $\beta$-\textit{smooth} on $Q\in \mathbb{R}^n$  with \textit{smooth constant} $\beta \ge 0$, if $f$ has $\beta$-\textit{Lipschitz continuous gradient}, i.e., for all  $\bm{x}, \bm{y} \in Q$, we have
\begin{equation}\label{smoothcondition}
\Vert \nabla f(\bm{x}) - \nabla f(\bm{y}) \Vert \le \beta \Vert \bm{x} - \bm{y} \Vert
\end{equation}
\end{definition}

\begin{lemma} \label{sumlipschitz}
The sum of smooth functions is smooth. Specifically, if a function $f:X\rightarrow Y$ is $\beta_f$-smooth on $X$, and a function $g: X\rightarrow Y$ is $\beta_g$-smooth on $X$, then the function $(f+g):X\rightarrow Z$ is $(\beta_f+\beta_g)$-smooth on $X$.
\end{lemma}

\subsection{Smooth Utility Function}

According to the analysis at the end of Section~\ref{gdsec} and Lemma~\ref{boundedgradient}, the $\alpha$-fair function is not Lipschitz continuous. Its corresponding network utility function  $U(\bm{x}) = \sum_{s\in S}U_{\alpha}(x_s)$ is not smooth, either. $\Vert \nabla U(\bm{x}) - \nabla  U(\bm{y}) \Vert = \sqrt{\sum_{s\in S} (x_s^{-\alpha} - y_s^{-\alpha})^2}$. If  $x_s = 0$ for some $s\in S$, then $x_s^{-\alpha} = \infty$ and $ \Vert \nabla U(\bm{x}) - \nabla U(\bm{y}) \Vert = \infty$. Thus, Eq.~\eqref{smoothcondition} does not hold. Clearly, the ``infinity'' problem occurs when some $x_s$ approaches $0$. So we define the $(\alpha, \xi)$-fair function as:
\begin{equation}
U_{\alpha, \xi}(x_s)=\left\{
\begin{array}{rcl}
&\!\!\!\!\!\!\!\!\dfrac{(x_s+\xi)^{1-\alpha}}{1-\alpha},       & {\alpha \ge0, \alpha \neq 1}\\
&\!\!\!\!\!\!\!\!\ln{(x_s+\xi)},  & {\alpha=1}\\
\end{array} \right.,\label{uutil2}
\end{equation}
where $\xi$ is a small number and $\xi>0$.
Because the formulation of the $(\alpha, \xi)$-fair function is similar to the $\alpha$-fair function, so we think it can be used as a link utility function. Actually, it is conventional to add a small number to the argument $x_s$ of the $\alpha$-fair function to avoid numerical issues when $x_s \rightarrow 0$. For example, Exp-NUM\cite{8737600} and Primal-dual\cite{6756941} did so in their experiments. However, existing works view the function $U_{\alpha, \xi}(x_s)$ as a ``mathematical trick'' instead of using it in problem formulation, and there is no theoretical analysis of the  mathematical property of $U_{\alpha, \xi}(x_s)$. Here we show that it yields a smooth network utility function.

\begin{lemma}\label{smoothutilty}
The function $U(\bm{x}) =\sum_{s\in S}U_{\alpha, \xi}(x_s) $ is  $\tfrac{\alpha}{\xi^{\alpha+1}}$-smooth on $\mathbb{R}_+^d$.
\end{lemma}
\begin{proof}
First we define function $\varphi: \mathbb{R}_+ \rightarrow \mathbb{R}_+$ as:
\begin{equation}
\varphi(x) = (x + \xi)^{-\alpha}, \nonumber
\end{equation}
where $\alpha \ge 0, \xi>0$. Then we have:
\begin{equation}
\left|\varphi'(x)\right| = \left|\frac{\alpha}{(x+\xi)^{\alpha+1}}\right| \le \frac{\alpha}{\xi^{\alpha+1}},\,\,\forall x\ge0.\nonumber
\end{equation}
By Lemma~\ref{boundedgradient}, $\varphi(x) $ is$\frac{\alpha}{\xi^{\alpha+1}}$-Lipschitz continuous. Then by Definition~\ref{lipschitz},
\begin{equation}
\left|\varphi(x)-\varphi(y)\right| \le \frac{\alpha}{\xi^{\alpha+1}} \left|x-y\right|,\,\,\forall x\ge0, y\ge0.\label{phi}
\end{equation}
By Eq.~\eqref{uutil2},
\begin{equation}
\nabla U\left(\bm{x}\right) = \nabla\left[\sum\nolimits_{s\in S}U_{\alpha,\xi}\left(\bm{x}\right) \right] = \left((x_s+\xi)^{-\alpha}\right)_{s\in S}. \label{grad}
\end{equation}
Then for all  $\bm{x}, \bm{y} \in \mathbb{R}_+^d$, we have:
\begin{align}
&\Vert \nabla U(\bm{x}) - \nabla  U(\bm{y}) \Vert \nonumber \\
= \,\,\,&\sqrt{\sum_{s\in S} \left[(x_s+\xi)^{-\alpha} - (y_s+\xi)^{-\alpha}\right]^2}  &\quad& \text{(by Eq.~\eqref{grad})}\nonumber \\
\le \,\,\,& \sqrt{\sum_{s\in S} \left[ \tfrac{\alpha}{\xi^{\alpha+1}} \left|x_s - y_s\right|\right]^2}  &\quad& \text{(by Eq.~\eqref{phi})}\nonumber \\
=\,\,\, & \tfrac{\alpha}{\xi^{\alpha+1}} \Vert \bm{x} - \bm{y}\nonumber  \Vert.
\end{align}
By Definition~\ref{smooth}, $U(\bm{x}) = \sum_{s\in S}U_{\alpha, \xi}(x_s)$ is $\tfrac{\alpha}{\xi^{\alpha+1}}$-smooth.
\end{proof}

\begin{remark}
A special case is when $\alpha = 0$. In this case, the function $ U(\bm{x}) = \sum_{s\in S}U_{0}(x_s) =  \sum_{s\in S}x_s$ is already $0$-smooth, so there is no need to use $U_{0, \xi}(x_s)$. In addition,  $\sum_{s\in S}U_{0, \xi}(x_s)$ is also $0$-smooth for any $\xi \in \mathbb{R}$. Thus, Lemma~\ref{smoothutilty} still holds for $\alpha = 0$. The proof is omitted due to the space limitation.
\end{remark}

\subsection{Smooth Penalty Function}

\begin{figure}[t]
\centerline{\includegraphics[width = 0.7\columnwidth]{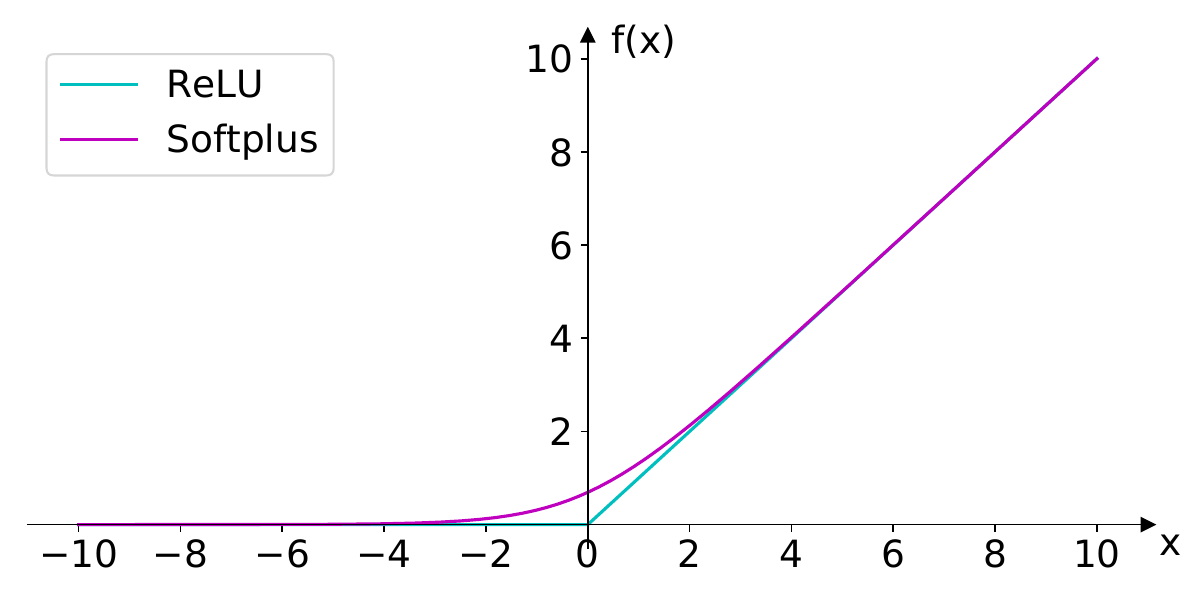}}
\caption{Graphs of ReLU function and Softplus function.}
\label{ss}
\end{figure}

The exact penalty function $P_{e,+}$ is not continuously differential and not smooth. Neither does the corresponding network penalty function. There have been works about approximating $P_{e,+}$ with functions with more tractable mathematical properties, see \cite{doi:10.1137/0804027, xu2013second, ZENIOS1995220} for example. Usually piecewise functions are used which are difficult to handle, and the functions are continuously differential but the smoothness is not proved.

We notice that the exact penalty function is actually the so called \textit{Rectified Linear Unit (ReLU)} activation function widely used in deep learning. Inspired by it, we propose to use the \textit{Softplus} function defined as
\begin{equation}
f(x) = \ln(1+\exp(x)), \,\, \forall x\in\mathbb{R},\label{softplus}\nonumber
\end{equation}
to construct the penalty function. Fig.~\ref{ss} shows the graphs of ReLU function and Softplus function. The two functions behavior similar except near the origin, where the Softplus function is differential and seems to be smooth.  Then, for $e\in E$ we define the soft penalty function $P_{e,S}$ as:
\begin{equation}
P_{e,S}(\mathbf{A}_e\bm{x}) = \ln(1+\exp({\mathbf{A}_e\bm{x}-c_e})).\nonumber
\end{equation}
And we then prove that its corresponding network penalty function is smooth.

\begin{lemma}\label{smoothpenalty}
The function $P(\bm{x})=\sum_{e\in E}P_{e,S}(\mathbf{A}_e\bm{x})$  is $(\frac{1}{4}\sum_{e\in E}\Vert \mathbf{A}_e\Vert^2)$-smooth on $\mathbb{R}_+^d$.
\end{lemma}
\vspace{-3mm}
\begin{proof}
For all $e\in E$, we define function $f_e:  \mathbb{R}_+\rightarrow \mathbb{R}_+^d $ as
\begin{equation}
f_e(x) = x\mathbf{A}_e,\nonumber
\end{equation}
which is $\Vert \mathbf{A}_e\Vert $-Lipschitz on $\mathbb{R}_+ $ by Definition~\ref{lipschitz}, because for all $x,y\in \mathbb{R}_+$, we have
\vspace{-1mm}
\begin{equation}
\Vert  f_e(x) -f_e(y) \Vert = \Vert (x-y) \mathbf{A}_e\Vert = \Vert \mathbf{A}_e \Vert \left|x-y\right|.\vspace{-1mm}\nonumber
\end{equation}
We define function $g: \mathbb{R} \rightarrow \mathbb{R}^+$ as
\vspace{-2mm}
\begin{equation}
g(x) = f'(x) = \frac{1}{1+e^{-x}},\vspace{-1mm}\nonumber
\end{equation}
which is $\frac{1}{4}$-Lipschitz by Lemma~\ref{boundedgradient}, because its derivative satisfies
\vspace{-2mm}
\begin{equation}
\left|g'(x)\right| = \left|\frac{e^{-x}}{(1+e^{-x})^2}\right| \le \frac{1}{4}, \,\, \forall x\in \mathbb{R}.\vspace{-1mm}\nonumber
\end{equation}
For all $e\in E$, we define function $h_e: \mathbb{R}_+^d \rightarrow \mathbb{R}$ as
\begin{equation}
h_e(\bm{x}) = \mathbf{A}_e\bm{x}-c_e,\nonumber
\end{equation}
which is $\Vert \mathbf{A}_e\Vert $-Lipschitz on $\mathbb{R}_+^d $ by Definition~\ref{lipschitz}, because
for all $\bm{x},\bm{y} \in \mathbb{R}_+^d$, we have
\begin{equation}
\Vert h_e(\bm{x})-h_e(\bm{y})\Vert = \left| \mathbf{A}_e\bm{x} - \mathbf{A}_e\bm{y} \right|\le \Vert \mathbf{A}_e\Vert  \Vert \bm{x}-\bm{y}\Vert.\nonumber
\end{equation}
For all $e\in E$, we define function $\tilde{P}_{e, S}(\bm{x})= P_{e,S}(\mathbf{A}_e\bm{x})$, whose gradient is
\begin{equation}
\nabla \tilde{P}_{e, S}(\bm{x}) = \frac{1}{1+\exp(-\mathbf{A}_e\bm{x}+c_e)}\mathbf{A}_e = f_e \circ g \circ h_e(\bm{x}).\nonumber
\end{equation}
By Lemma~\ref{compositionlipschitz}, $\nabla \tilde{P}_{e, S}(\bm{x})$ is $(\frac{1}{4}\Vert \mathbf{A}_e\Vert^2)$-Lipschitz on $\mathbb{R}_+^d$, i.e., $\tilde{P}_{e, S}(\bm{x})$ is $(\frac{1}{4}\Vert \mathbf{A}_e\Vert^2)$-smooth on $\mathbb{R}_+^d$. Then by Lemma~\ref{sumlipschitz}, the function $P(\bm{x})=\sum_{e\in E}P_{e,S}(\mathbf{A}_e\bm{x})$  is $(\frac{1}{4}\sum_{e\in E}\Vert \mathbf{A}_e\Vert^2)$-smooth on $\mathbb{R}_+^d$.
\end{proof}

\subsection{NUM with Smooth Objective Function}

With the smooth utility function and smooth penalty function, the NUM problem is now defined as:
\begin{equation}\label{nums}\tag{NUM-S}
  \begin{aligned}
    \min \,\,\,V_S(\bm{x})=&-{\sum_{s \in S}U_{\alpha, \xi}(x_s)}+\mu\sum_{e\in E}P_{e,S}(\mathbf{A}_e \bm{x})\\
  \text{s.t.,}\,\,\,\, & x_s \ge 0,\,\, \forall s\in S
  \end{aligned}\!\!\!\!\!\!\!
\end{equation}

\begin{theorem}\label{smoothv}
The function $V_S$ is $\beta_V$-smooth on $\mathbb{R}_+^d$, where  $\beta_V = \tfrac{\alpha}{\xi^{\alpha+1}} + \frac{\mu}{4}\sum_{e\in E}\Vert \mathbf{A}_e\Vert^2=O(d)$ .
\end{theorem}
\begin{proof}
We define $U(\bm{x})\!\!\!\!=\!\!\!\!\sum_{s\in S}U_{\alpha, \xi}(x_s)$  and  $P(\bm{x})\!\!=\!\!\sum_{e\in E}P_{e,S}(\mathbf{A}_e\bm{x})$, then
\begin{equation}
\nabla V_S(\bm{x}) = - \nabla U(\bm{x}) + \mu \nabla P(\bm{x}).\nonumber
\end{equation}
Then for all $\bm{x}, \bm{y} \in \mathbb{R}_+^d$, we have
\begin{align}
&\Vert \nabla V_S(\bm{x}) - \nabla  V_S(\bm{y}) \Vert \nonumber \\
=\,\,&\Vert \nabla U(\bm{y})  - \nabla U(\bm{x})   +  \mu\nabla P(\bm{x}) - \mu\nabla P(\bm{y})\Vert  \nonumber\\
\le \,\,&\Vert \nabla U(\bm{y})  - \nabla U(\bm{x}) \Vert +\mu \Vert \nabla P(\bm{x}) - \nabla P(\bm{y})\Vert \nonumber\\
\le\,\, & (\tfrac{\alpha}{\xi^{\alpha+1}} + \tfrac{\mu}{4}\textstyle\sum_{e\in E}\Vert \mathbf{A}_e\Vert^2 ) \Vert \bm{x} - \bm{y}\Vert &\!\!\!\!\!\!\!\!\!\!\!\!\!\!\!\!\!\!\!\!\!\!\!\! \text{(by Lemma \ref{smoothutilty}, \ref{smoothpenalty})}\nonumber
\end{align}
By Definition~\ref{smooth}, the function $V_S$ is  $(\tfrac{\alpha}{\xi^{\alpha+1}} + \frac{\mu}{4}\sum_{e\in E}\Vert \mathbf{A}_e\Vert^2)$-smooth on $\mathbb{R}_+^d$. Besides, by definition of the routing matrix $\mathbf{A}$ (Eq.~\eqref{A}), we have:
\begin{equation}
\beta_V = \frac{\alpha}{\xi^{\alpha+1}} + \frac{\mu}{4}\sum_{e\in E}\Vert \mathbf{A}_e\Vert^2 \le \frac{\alpha}{\xi^{\alpha+1}} + \frac{\mu}{4}\left|E\right|d.\label{eq9}
\end{equation}
\ggai{By Assumption~\ref{ass_E} and Assumption~\ref{ass_D}, $|E|\ll d$.} Thus, by Eq.~\eqref{eq9}, $\beta_V = O(d)$.
\end{proof}

One benefit of \eqref{nums} is that Assumption~\ref{ass_L} holds for its objective function $V_S$. For all $\bm{x}\in \mathbb{R}_+^d, s \in S$, $\left|\frac{\partial V_p(\bm{x})}{\partial x_s}\right| \le M = \max\{\xi^{-\alpha}, \mu\sum_{e\in E}\mathbf{A}_{es}\;\forall s\in S\}$.
 Thus, PGD and Exp-NUM can actually be used to solve \eqref{nums} \rev{and the converge analyses become solid}.
Furthermore, because the objective function is now smooth, the convergence guarantee of PGD can be improved as follows:
\begin{proposition}[Convergence rate of PGD for \eqref{nums}\cite{MAL-050}] \label{prop_pgd2}
If Algorithm~\ref{pgd} is run for $T$ iterations with an initial point $\bm{x}^{(0)}$, step size $\gamma = \tfrac{1}{\beta_V}$, then the solution  $\bm{x}^{(T)}$  satisfies:
\begin{equation}
V_S(\bm{x}^{(T)}) - V_S^{*} \le \frac{2\beta_V\Vert \bm{x}^{*}\!\!-\bm{x}^{(0)} \Vert^2}{T}.\label{pgdconvergeeq}
\end{equation}
By Theorem~\ref{smoothv}, $\beta_V = O(d)$. Thus, the error is $\varepsilon= O(d/t)$ at step $t$, or equivalently, PGD achieves an $\varepsilon$-optimal solution with $\Omega(d/\varepsilon)$ iterations.
\end{proposition}
For \eqref{nums}, the convergence rate of PGD is now proportional to $1/t$ instead of $1/\sqrt{t}$, but the iteration complexity is still proportional to $d$. Note that now the final solution is not the averaged point, but the point at the last iteration.
However, Exp-NUM is only applicable to minimizing Lipschitz functions, so its convergence rate remains $O(\sqrt{\ln{d}}/\sqrt{t})$ for \eqref{nums}.


\section{Accelerated Gradient Method for NUM with Smooth Objection Function}\label{agmsec}

In this section, we present the accelerated gradient method (AGM), which is a tailor-made algorithm for smooth objective function.

\subsection{Vanilla Accelerated Gradient Method}

\begin{algorithm}[tb]
\renewcommand{\algorithmicrequire}{\textbf{Input:}}
\renewcommand{\algorithmicensure}{\textbf{Output:}}
\caption{Accelerated Gradient Method (AGM)}\label{agm}
\begin{algorithmic}[1]
\REQUIRE  initial point $\bm{x}^{(0)}$, step size $\gamma$, number of iterations $T$
\ENSURE final solution $\bm{x}^{(T)}$
\STATE $\bm{y}^{(0)} \leftarrow \bm{x}^{(0)}$, $a_0 \leftarrow 1$
\FOR{$t \leftarrow 0$ \textbf{to} ${T-1}$}
    \STATE $\bm{x}^{(t+1)} \leftarrow \left[\bm{y}^{(t)} - \gamma \nabla V_S(\bm{y}^{(t)})\right]_+$
    \STATE $a_{t+1} \leftarrow \dfrac{1+\sqrt{4a_t^2+1}}{2} $
    \STATE$\bm{y}^{(t+1)} \leftarrow \bm{x}^{(t+1)} + \dfrac{a_t-1}{a_{t+1}}(\bm{x}^{(t+1)}-\bm{x}^{(t)})$
\ENDFOR
\RETURN  $\bm{x}^{(T)} $
\end{algorithmic}
\end{algorithm}

We have shown that the proposed \eqref{nums} problem benefits the use of PGD and Exp-NUM methods. Furthermore, it also enables the use of the \textit{accelerated gradient method} (AGM) proposed by Nesterov\cite{nesterov1983method}, which achieves a superior $O(1/t^2)$ convergence rate for smooth objective functions.

The pseudocode of AGM is shown in Algorithm~\ref{agm}. At each iteration, $\bm{x}^{(t+1)}$ is moved towards the opposite gradient direction at the point $\bm{y}^{(t)}$ instead of $\bm{x}^{(t)}$. Then the point $\bm{y}^{(t+1)}$ is updated based on the obtained point $\bm{x}^{(t+1)}$ at this iteration and $\bm{x}^{(t)}$ at previous iteration, balanced by the parameters $a_t$ and $a_{t+1}$. AGM can be viewed as a momentum method\cite{pmlr-v28-sutskever13},  because it takes steps based on information at not only the current iteration but also previous iterations. \rev{It} has the following convergence guarantee:

\begin{theorem}[Convergence rate of AGM\cite{nesterov1983method}]\label{agdconverge}
If Algorithm~\ref{agm} is run for $T$ iterations with an initial point $\bm{x}^{(0)}$, step size $\gamma = \tfrac{1}{\beta_V}$, then the final solution  $\bm{x}^{(T)}$  satisfies:
\begin{equation}
V_S(\bm{x}^{(T)}) - V_S^{*} \le \frac{{2}\beta_V\Vert \bm{x}^{*}\!\!-\bm{x}^{(0)} \Vert^2}{(T+1)^2}.\label{agdconvergeeq}
\end{equation}
By Theorem~\ref{smoothv}, $\beta_V = O(d)$. Thus, the error is $\varepsilon= O(d/t^2)$ at step $t$, or equivalently, AGM achieves an $\varepsilon$-optimal solution with $\Omega(\sqrt{d}/\sqrt{\varepsilon})$ iterations.
\end{theorem}

Like PGD and Exp-NUM, AGM is also a first-order method with low per-iteration cost. In addition, AGM has two main advantages over PGD and Exp-NUM:

1) AGM has the fastest convergence rate with respect to $t$, which is proportional to $1/t^2$. Compared with PGD (convergence rate proportional to $1/t$) and Exp-NUM (convergence rate proportional to $1/\sqrt{t}$), AGM \rev{is much more efficient, and it} converges to an accurate solution within a lot less iterations.

2) The iteration complexity of AGM is proportional to $\sqrt{d}$. Although AGM is not as scalable as Exp-NUM with respect to the number of flows, it still can maintain good performance as the number of flows grows.



\begin{remark}
\ggai{The convergence guarantees of PGD, Exp-NUM and AGM hold under different assumptions: Assumption~\ref{ass_L} or~\ref{ass_U} for  PGD (Proposition~\ref{prop_gd} or~\ref{prop_pgd2}), Assumption~\ref{ass_L}-\ref{ass_C} for Exp-NUM (Proposition~\ref{prop_expnum}),  and Assumption~\ref{ass_E}-\ref{ass_U} for AGM (Theorem~\ref{smoothv},~\ref{agdconverge}).  When Assumption \ref{ass_E}-\ref{ass_C} are all true, Proposition \ref{prop_expnum}, \ref{prop_pgd2} and Theorem \ref{agdconverge} all hold, and} it is provable that AGM achieves better solution compared with PGD throughout the whole iteration process, and AGM guarantees to achieve more accurate solution compared with Exp-NUM after enough number of iterations.
\end{remark}

\begin{theorem}\label{agd_vs_pgd}
\rev{The solution achieved by AGM is better than that achieved by PGD for any number of iterations $T$.}
\end{theorem}
\begin{proof}
\rev{By Eq.~\eqref{pgdconvergeeq} and Eq.~\eqref{agdconvergeeq}, for AGM to achieve better solution, there should be:
\begin{equation}
\frac{{2}\beta_V\Vert \bm{x}^{*}\!\!-\bm{x}^{(0)} \Vert^2}{(T+1)^2} \le \frac{2\beta_V\Vert \bm{x}^{*}\!\!-\bm{x}^{(0)} \Vert^2}{T},\nonumber
\end{equation}
\vspace{-3mm}
\begin{equation}
\Longrightarrow\quad\frac{(T+1)^2}{T} \ge 1.\label{teq1}
\end{equation}
Obviously, Eq.~\eqref{teq1} holds for any $T\ge1$.}
\end{proof}

\begin{theorem}\label{agd_vs_expnum}
\rev{The solution achieved by AGM is better than that achieved by Exp-NUM when the number of iterations is larger than  $\left(\frac{2\beta_{V}\Vert \bm{x}^{*}\!\!-\bm{x}^{(0)} \Vert^4}{M^2CD(\bm{x}^{*}, \bm{x}^{(0)})}\right)^{\tfrac{1}{3}}-1$.}
\end{theorem}
\begin{proof}
\rev{By Eq.~\eqref{expnumconvergeeq} and Eq.~\eqref{agdconvergeeq}, for AGM to achieve better solution, there should be:
\begin{equation}
\frac{{2}\beta_V\Vert \bm{x}^{*}\!\!-\bm{x}^{(0)} \Vert^2}{(T+1)^2} \le \frac{M\sqrt{2CD(\bm{x}^{*}, \bm{x}^{(0)})}}{\sqrt{T}},\nonumber
\end{equation}
\begin{equation}
\Longrightarrow\quad\frac{(T+1)^4}{T} \ge \frac{{2}{\beta_V}^2\Vert \bm{x}^{*}\!\!-\bm{x}^{(0)} \Vert^4}{M^2CD(\bm{x}^{*}, \bm{x}^{(0)})}.\label{teq2}
\end{equation}
As $\frac{(T+1)^4}{T} \ge \frac{(T+1)^4}{T+1}$, for Eq.~\eqref{teq2} to hold, it is sufficient that the following inequality holds:
\begin{equation}
\frac{(T+1)^4}{T+1} \ge \frac{{2}{\beta_V}^2\Vert \bm{x}^{*}\!\!-\bm{x}^{(0)} \Vert^4}{M^2CD(\bm{x}^{*}, \bm{x}^{(0)})}.\label{teq3}
\end{equation}
By Eq.~\eqref{teq3}, $T\ge \left(\dfrac{2\beta_{V}\Vert \bm{x}^{*}\!\!-\bm{x}^{(0)} \Vert^4}{M^2CD(\bm{x}^{*}, \bm{x}^{(0)})}\right)^{\tfrac{1}{3}}-1$.}
\end{proof}


\captionsetup[subfloat]{position=bottom}
\begin{figure*}[tb]
\begin{minipage}{\textwidth}
\footnotesize{
~~~~~~~~~~~~~~~~~~~~~~~~~~~~~~~~~~~~~~~~~~~{G\'EANT}~~~~~~~~~~~~~~~~~~~~~~~~~~~~~~~~~~~~~~~~~~~~~~~~~~~~~~~~~~~~~~~~~~~~~~~~~~~~~~~~~~~RF1221}

\footnotesize{~~~~~~~~~~~~~~~~~$U(\bm{x}) = \textstyle\sum_{s\in S}x_s$~~~~~~~~~~~~~~ $U(\bm{x}) = \sum_{s\in S}\ln(x_s+0.5)$~~~~~~~~~~~~~~~~~ $U(\bm{x}) = \sum_{s\in S}x_s$~~~~~~~~~~~~~~~ $U(\bm{x}) = \sum_{s\in S}\ln(x_s+0.5)$}\vspace{-3mm}
\end{minipage}
\begin{minipage}{\textwidth}
\centering
    \subfloat{\label{wa1}\includegraphics[width=0.25\textwidth]{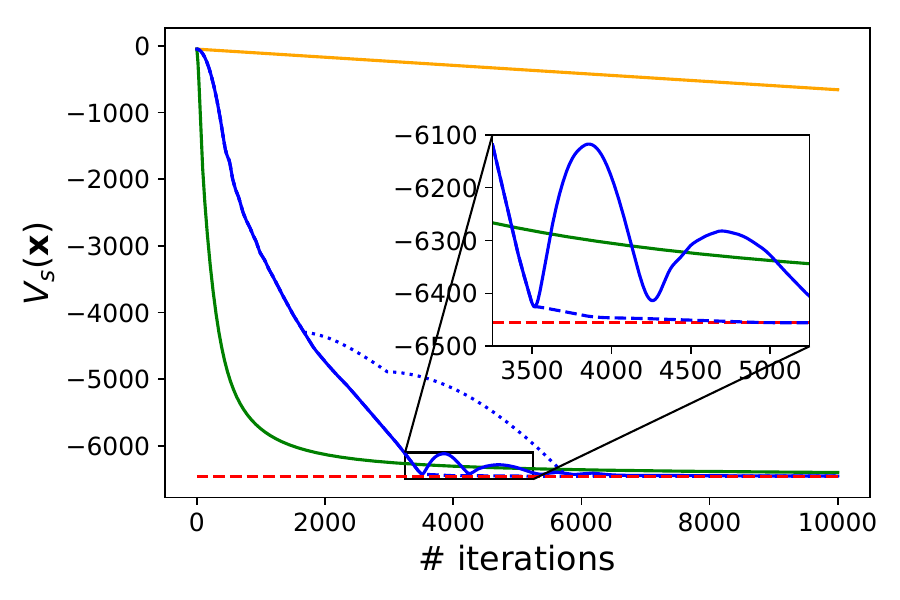}}
    \subfloat{\label{wa2}\includegraphics[width=0.25\textwidth]{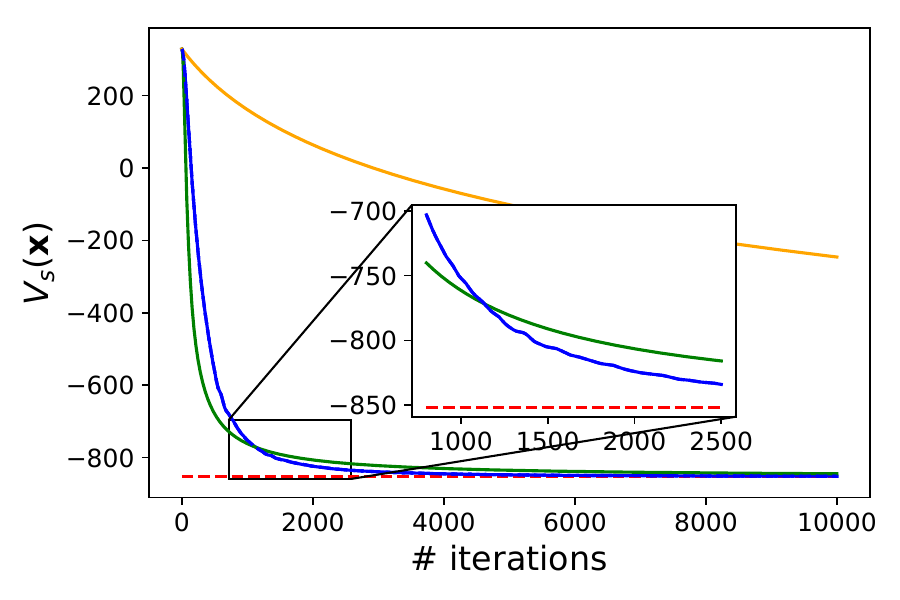}}
     \subfloat{\label{wa3}\includegraphics[width=0.25\textwidth]{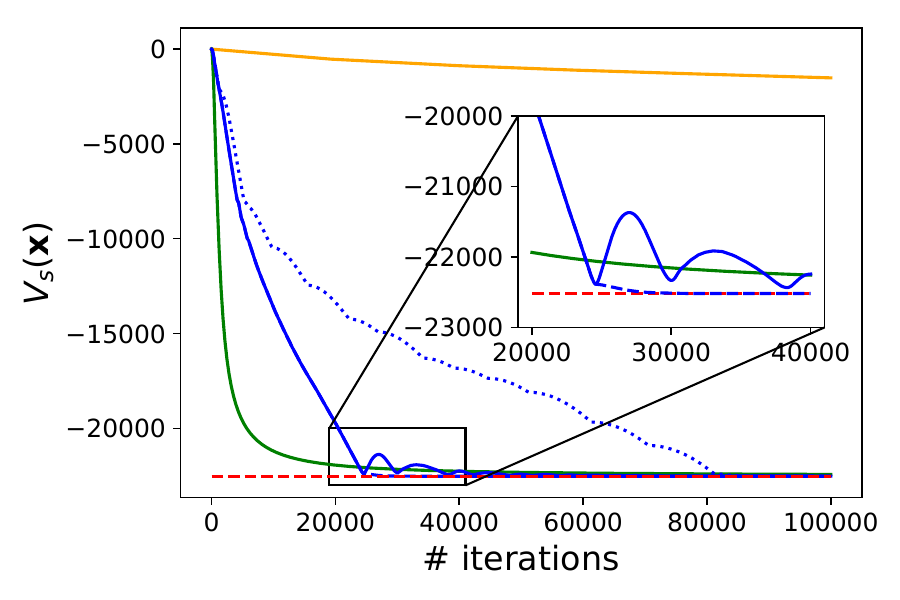}}
      \subfloat{\label{wa4}\includegraphics[width=0.25\textwidth]{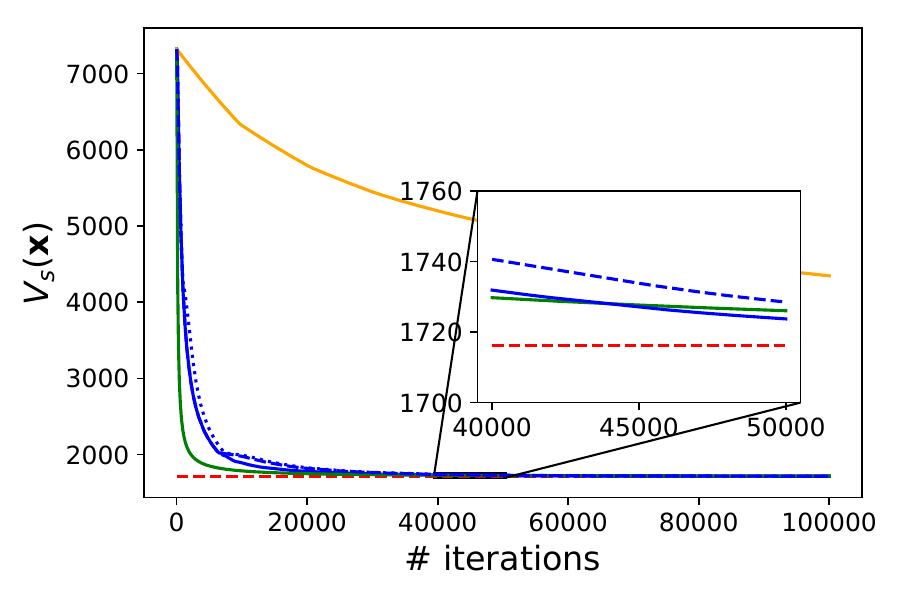}}\\\vspace{-3.5mm}
     \subfloat{\label{wa5}\includegraphics[width=0.25\textwidth]{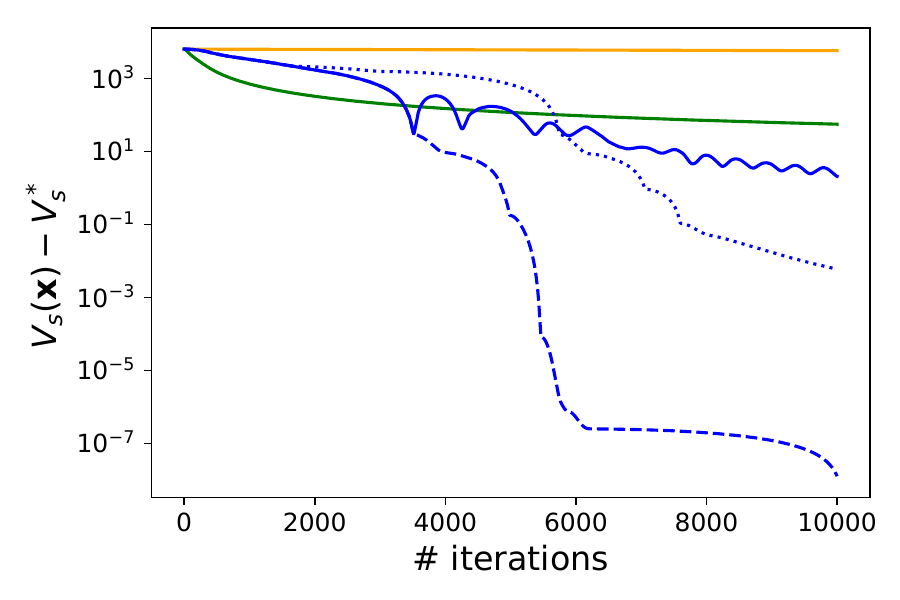}}
     \subfloat{\label{wa6}\includegraphics[width=0.25\textwidth]{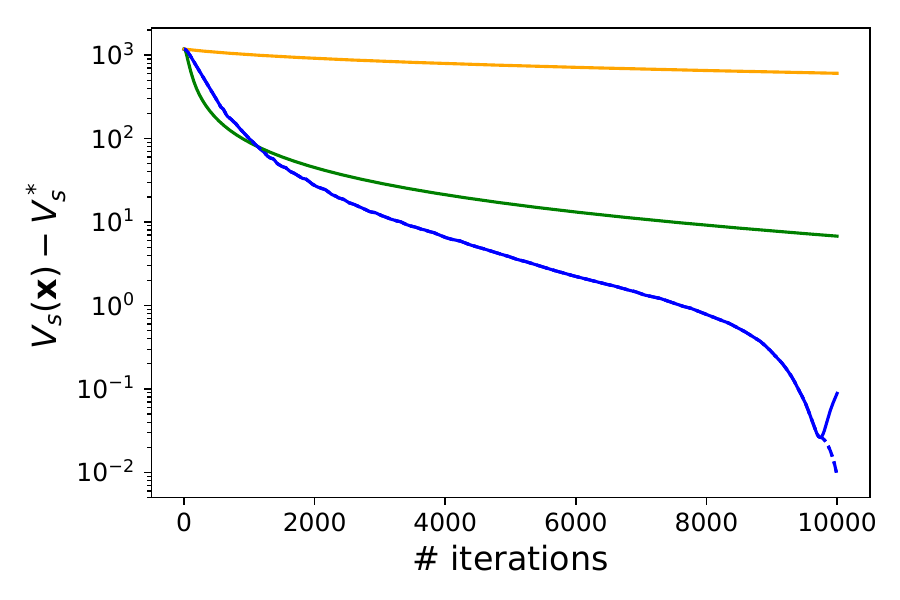}}
      \subfloat{\label{wa7}\includegraphics[width=0.25\textwidth]{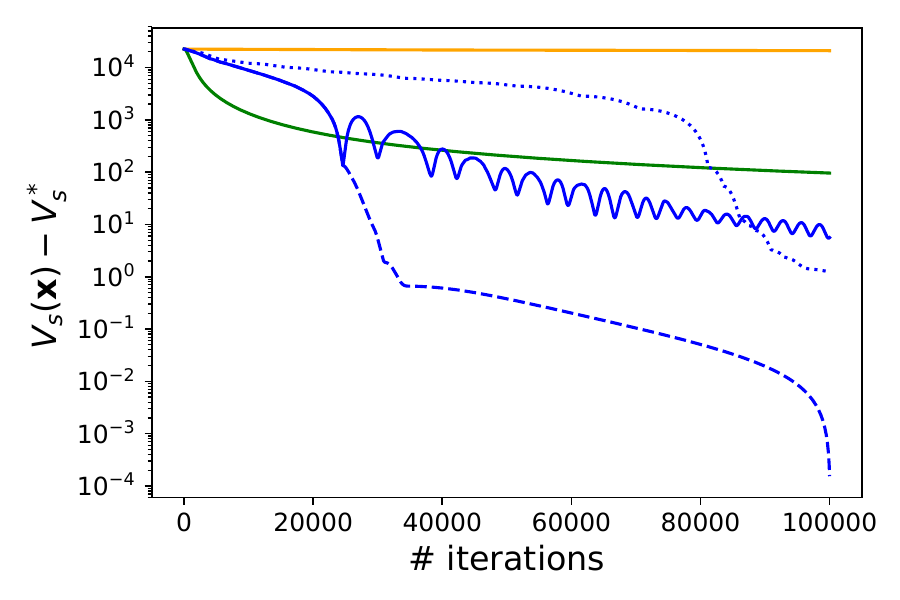}}
      \subfloat{\label{wa8}\includegraphics[width=0.25\textwidth]{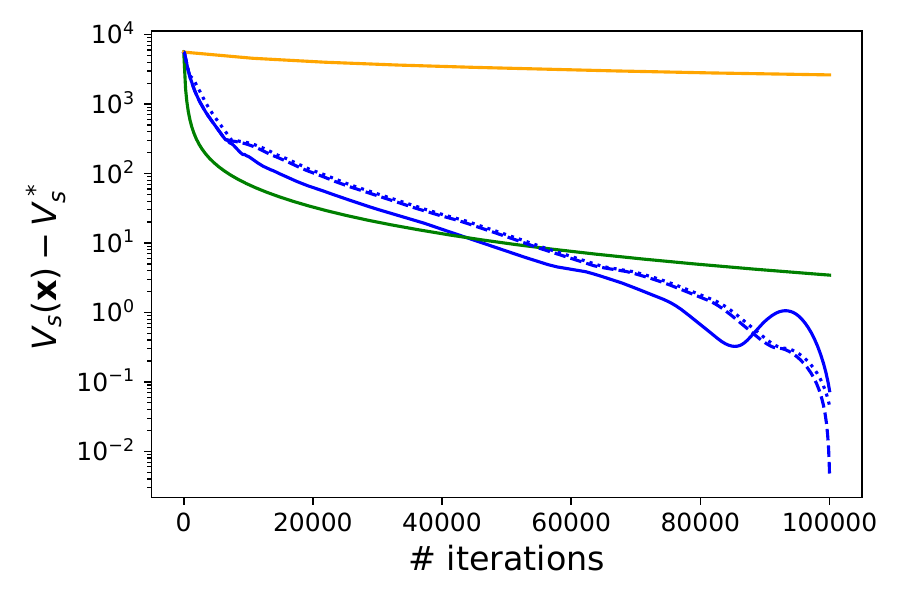}}

\end{minipage}
\\
\begin{minipage}{\textwidth}
\centerline{\includegraphics[width=0.45\textwidth]{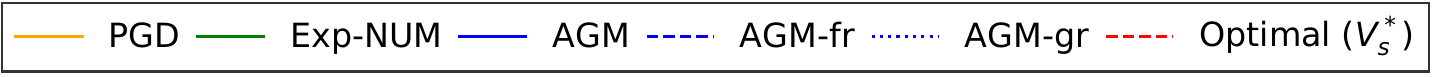}}
\end{minipage}
\caption{The objective value and error at each iteration in G\'EANT and RF1221.}\label{result1}
\vspace{-4mm}
\end{figure*}

\subsection{\vspace{3mm}Accelerated Gradient Method with Restart}

\begin{algorithm}[tb]
\renewcommand{\algorithmicrequire}{\textbf{Input:}}
\renewcommand{\algorithmicensure}{\textbf{Output:}}
\caption{Accelerated Gradient Method with Restart}\label{agm2}
\begin{algorithmic}[1]
\REQUIRE  initial point $\bm{x}^{(0)}$, step size $\gamma$, number of iterations $T$
\ENSURE final solution $\bm{x}^{(T)}$
\STATE $\bm{y}^{(0)} \leftarrow \bm{x}^{(0)}$, $a_0 \leftarrow 1$
\FOR{$t \leftarrow 0$ \textbf{to} ${T-1}$}
    \STATE $\bm{x}^{(t+1)} \leftarrow \left[\bm{y}^{(t)} - \gamma \nabla V_S(\bm{y}^{(t)})\right]_+$
    \STATE $a_{t+1} \leftarrow \dfrac{1+\sqrt{4a_t^2+1}}{2} $
    \STATE$\bm{y}^{(t+1)} \leftarrow \bm{x}^{(t+1)} + \dfrac{a_t-1}{a_{t+1}}(\bm{x}^{(t+1)}-\bm{x}^{(t)})$
     \IF{Restart Condition}
             \STATE $\bm{y}^{(t+1)} \leftarrow \bm{x}^{(t+1)}$, $a_{t+1} \leftarrow 0$
       \ENDIF
\ENDFOR
\RETURN  $\bm{x}^{(T)} $
\end{algorithmic}
\end{algorithm}

Although by Theorem~\ref{agdconverge} AGM converges fast, it does not mean that the obtained objective value sequence is monotone. When running the AGM algorithm,  it is common that there are bumps in the objective value trajectory, which empirically slows down the convergence. To deal with this problem, we leverage the restart heuristic proposed by O'donoghue et al\cite{aaaaa}. They propose to ``restart'' the algorithm, i.e., taking the current point as a new initial point, when certain restart condition holds. In this paper, we consider two restart conditions:

1) function restart: $V_S(\bm{x}^{(k+1)}) > V_S(\bm{x}^{(k)})$;

2) gradient restart: $\left[\nabla V_S(\bm{y}^{(k)})\right]^\intercal \!\! \left(\bm{x}^{(k+1)} - \bm{x}^{(k)}\right) > 0$.

The pseudocode of AGM with restart is shown in Algorithm~\ref{agm2}. At each iteration, the algorithm checks whether the restart condition holds. If so, $\bm{y}^{(t+1)}$ is set to $\bm{x}^{(t+1)}$ and the parameter $a_{t+1}$ is reset to $1$, so that the algorithm is restarted by the current point. Although the restart method boosts convergence empirically, there is no theoretical guarantee on the convergence rate.

%
%
%


\section{Evaluation}\label{evaluation}

In this section, we evaluate the smooth penalty-based NUM formulation and the proposed algorithm.

\subsection{Setup}

\textit{1) Topology:} 
We use two topologies G\'EANT (23 nodes, 74 links)\cite{uhlig2006providing} and RF1221 (104 nodes, 302 links)\cite{10.1145/2829988.2787495}. The shortest path is used as the predetermined path for each flow, \gai{so elements in the routing matrix are integer}. To better display the result, the link capacities are scaled down so that $c_e\le 100,\forall e\in E$. \gai{Because the flow rates $x_s$ are decision variables in the NUM problem, the traffic matrix is not needed in the experiments.}

\textit{2) Optimization Objective:}
 We use two different utility functions. One is $\alpha=0, \xi=0$, so that the network utility function is $U(\bm{x}) = \sum_{s\in S}x_s$, corresponding to throughput maximization. The other is $\alpha=1, \xi=0.5$, so that  network utility function is $U(\bm{x}) = \sum_{s\in S}\ln(x_s+0.5)$, corresponding to proportional fairness. $\mu$ is set to 2 empirically.

\textit{3) Algorithms:} We solve the problem~\eqref{nums} with PGD, Exp-NUM ($C$ is set to $\sum_{e\in E}c_e$), AGM, AGM with function restart (i.e., AGM-fr) and AGM with gradient restart (i.e., AGM-gr).

All algorithms are implemented with Python. Computations are performed on a server with 2.40GHz Intel CPU and 512G memory.\vspace{-2mm}

\subsection{\vspace{-1mm}Convergence Rate}\label{exp1}

We evaluate the convergence rate of the algorithms. We assume that there is one flow between each source-destination pairs, and run the algorithms with the aforementioned objective. In Fig.~\ref{result1}, we plot the objective value (at top) and corresponding error (at bottom in log scale) in G\'EANT and RF1221 at the first 10000 and 100000 iterations respectively.

First, experiments show that AGM has the fastest convergence rate and outperforms PGD and Exp-NUM, which agrees with the theoretical results. \gai{Given a tolerable error, our method will solve the NUM problem much quicker, which benefits the network resource allocation.} \rrev{As proved in Theorem~\ref{agd_vs_pgd},} PGD converges the slowest in experiment, mainly because its iteration complexity is $\Omega(d/\varepsilon)$, which is proportion to $d$.
\rrev{When comparing AGM and Exp-NUM, the experiment results is in consistent with Theorem~\ref{agd_vs_expnum}}. Exp-NUM has $O(\sqrt{\ln{d}}/\sqrt{t})$ convergence rate, which is nearly independent on the number of flows $d$, thus it converges fast in the beginning. However, the convergence slows down later and Exp-NUM can only achieve $\text{10}\sim\text{10}^\text{2}$ error at end. In contrast, AGM converges a little slower at first, but it performs better and obtains more accurate solutions as the number of iterations grows, because of its $O(d/t^2)$ convergence rate, which is the fastest with respect to $t$. 

Second, the restart method boosts the convergence of AGM and alleviates the bumps. Although AGM converges a little faster at first, AGM-fr and AGM-gr outperform it as the iterations continue. In particular, AGM-fr achieves $\text{10}^{\text{-2}}$-optimal solution for all objective functions and topologies, and it even achieves $\text{10}^{\text{-7}}$-optimal solution for $U_s=\sum_{s\in S}x_s$ in G\'EANT. Besides, there are less bumps in the curves of AGM-fr and AGM-gr.

\captionsetup[subfloat]{position=bottom}
\begin{figure}[tb]
\begin{minipage}{\columnwidth}
\footnotesize{
~~~~~~~~~~~~~~~~~~~{G\'EANT}~~~~~~~~~~~~~~~~~~~~~~~~~~~~~~~~~~~~~RF1221}
\vspace{-5mm}
\end{minipage}
\begin{minipage}{\columnwidth}
\centering
    \subfloat{\label{a1}\includegraphics[width=0.5\columnwidth]{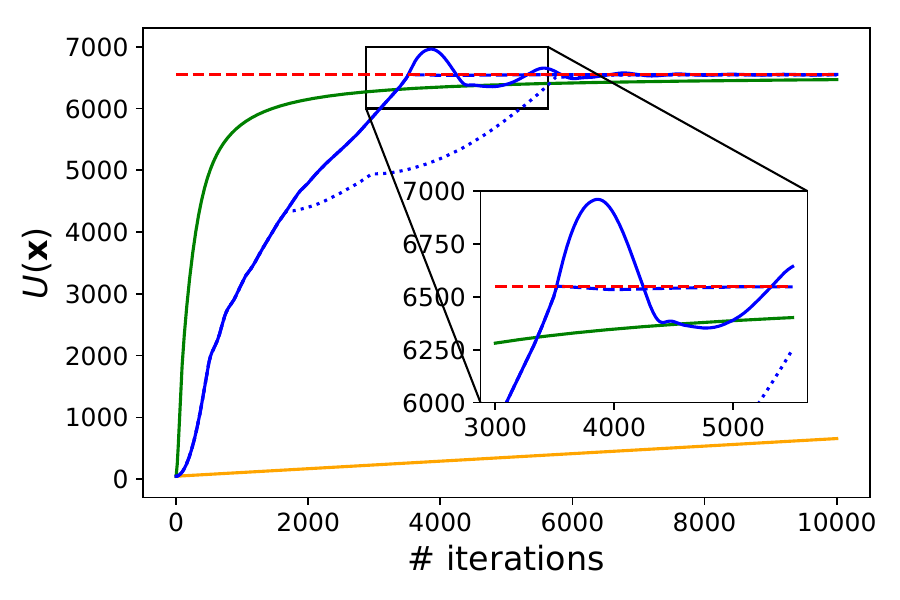}}
      \subfloat{\label{a4}\includegraphics[width=0.5\columnwidth]{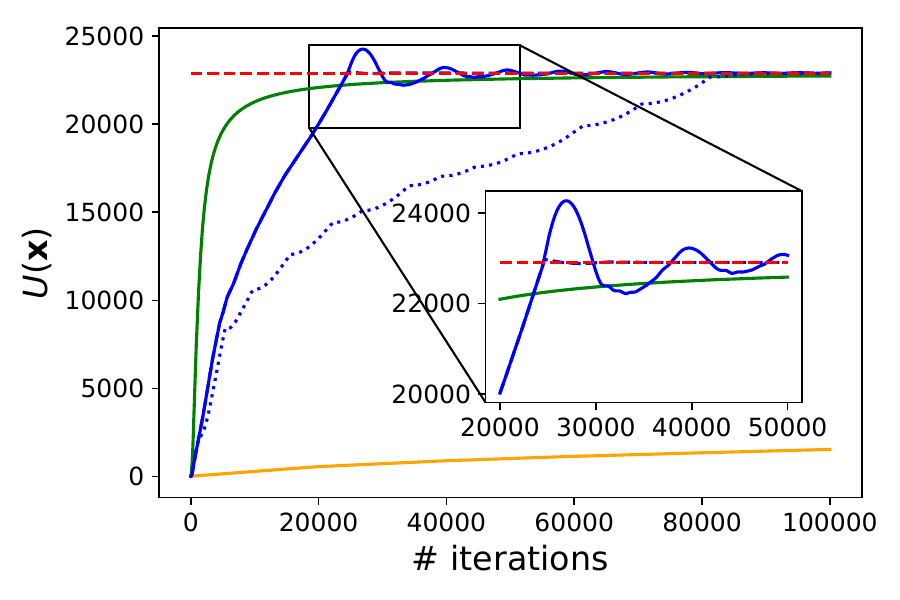}}\\\vspace{-3.5mm}
      \subfloat{\label{aa}\includegraphics[width=0.5\columnwidth]{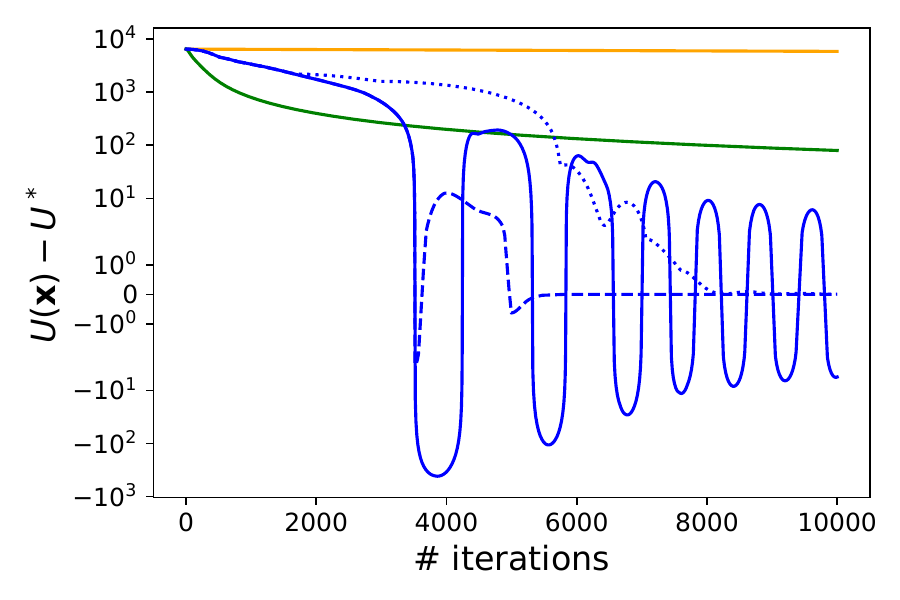}}
      \subfloat{\label{a21}\includegraphics[width=0.5\columnwidth]{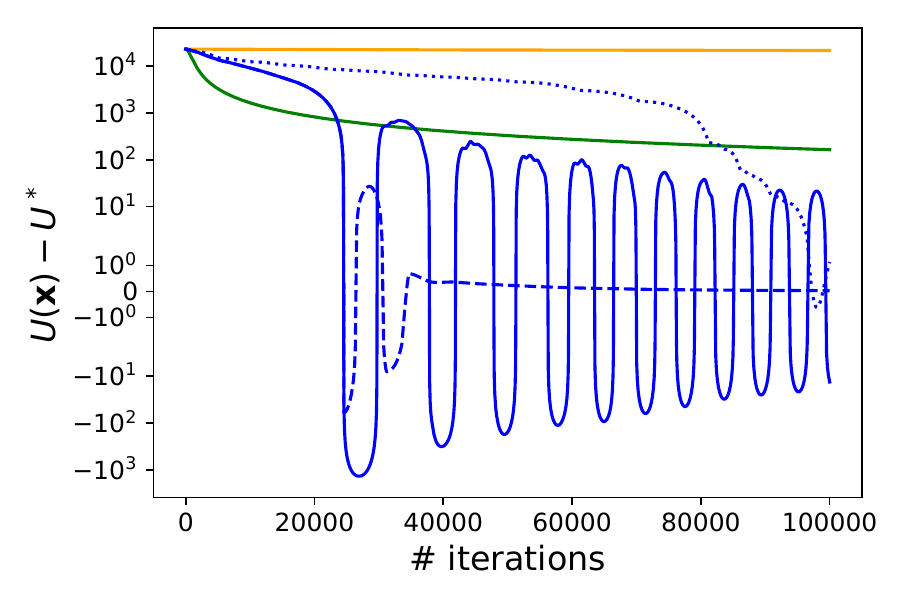}}

\end{minipage}
\\
\begin{minipage}{\columnwidth}
\centerline{\includegraphics[width=0.85\columnwidth]{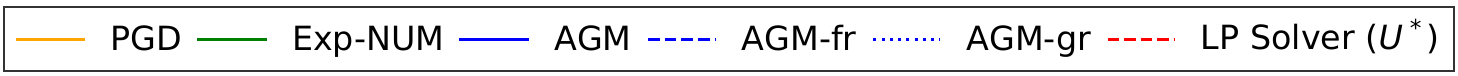}}
\end{minipage}
\caption{The network utility value and error at each iteration when $U_s=\sum_{s\in S}x_s$ in G\'EANT and RF1221.}\label{resultutil}
\end{figure}

\begin{figure*}[tb]
  \centering
  \captionsetup[subfloat]{captionskip=2pt} %
    \captionsetup[figure*]{captionskip=2pt} %
    \subfloat[$d~=$ 100]{\label{util1}\includegraphics[width=0.23\textwidth]{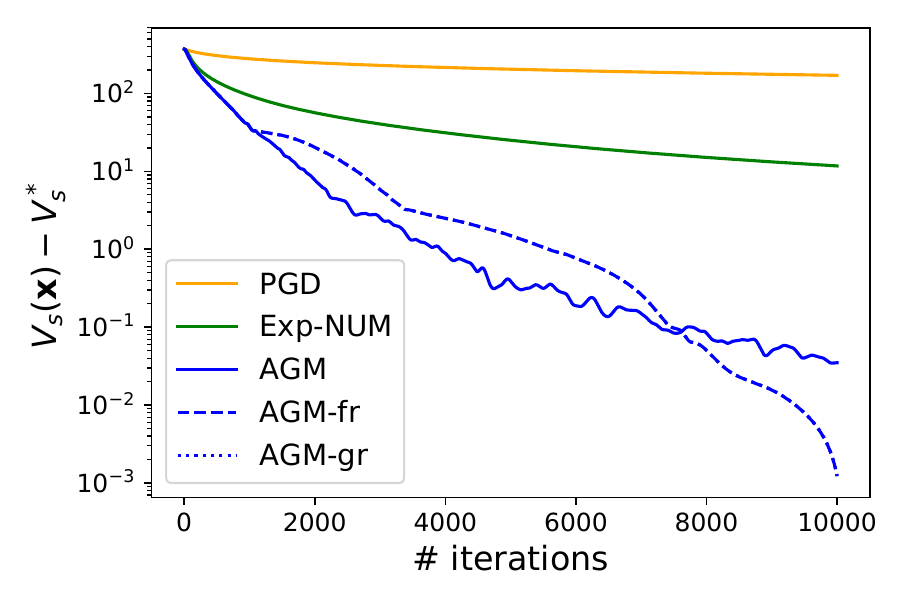}}
    \subfloat[$d~=$ 1000]{\label{util2}\includegraphics[width=0.23\textwidth]{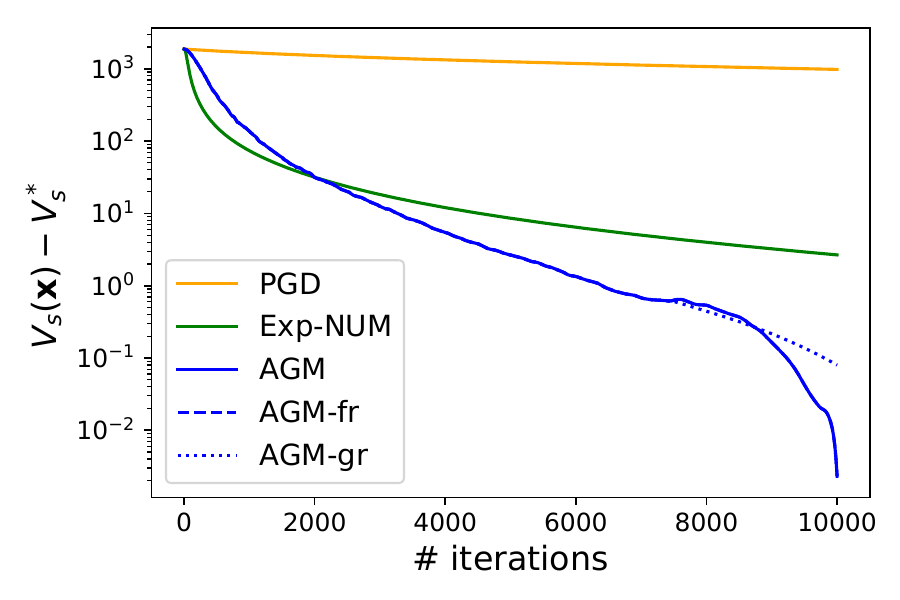}}
    \subfloat[$d~=$ 10000]{\label{util3}\includegraphics[width=0.23\textwidth]{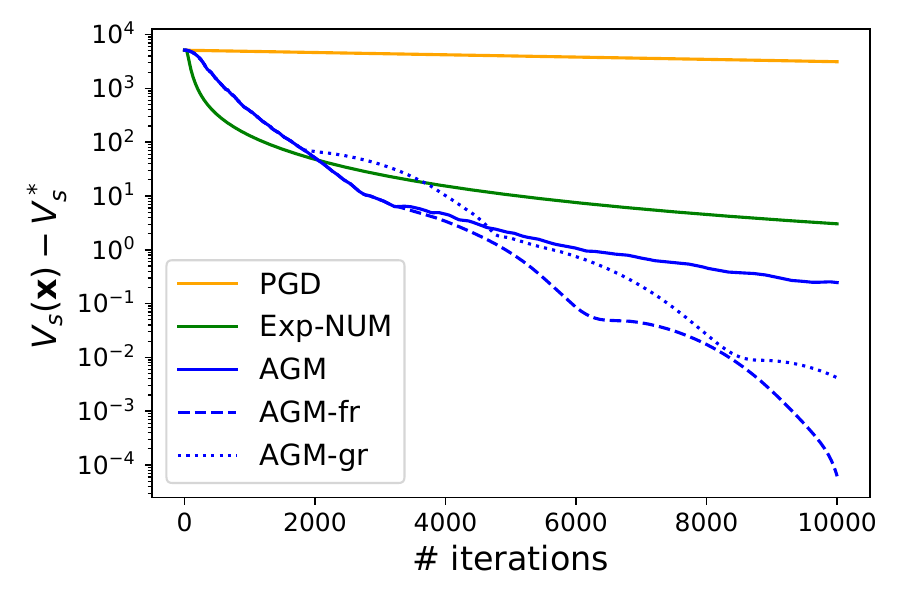}}
  \caption{The error at each iteration with different number of flows $d$ when $U_s=\sum_{s\in S}\ln{(x_s+0.5)}$ in G\'EANT.}
  \label{result2}
  \vspace{-5mm}
\end{figure*}

\subsection{\vspace{-1mm}Network Utility Optimization}\label{exp2}

The objective of \eqref{nums} is the smooth penalized function $V_s(\bm{x})$, instead of the network utility function $U(\bm{x})$. In order to evaluate the performance of the algorithms in terms of  network utility maximization, we take $U(\bm{x})=\sum_{s\in S}x_s$ for example so that the problem \eqref{num} is a linear programming (LP) problem aiming at maximizing throughput and can be solved by commercial LP solvers. We denote its optimal value as $U^*$. And we record the $U(\bm{x})$ (at top) and error with respect to $U^*$ (at bottom in symmetric log scale) during iterations when solving \eqref{nums}. The results are shown in Fig.~\ref{resultutil}.

First, the rate at which $U(\bm{x})$ approaches $U^*$ is consistent with the convergence rate of the algorithms, i.e., Exp-NUM performs well at first but AGM surpasses it later and PGD hardly converges. Hence, the convergence rate implies the efficiency of the algorithms in solving NUM problem.

Second, AGM causes obvious bumps in $U(\bm{x})$. Although it converges faster than Exp-NUM, it tends to overshoot the solution and oscillates around $U^*$. In contrast, AGM-fr obtains stable results while maintaining the fast convergence rate.

Third, AGM-fr achieves close to optimal network utility value, which means that the smooth penalty-based formulation and our algorithm can help solve the original NUM problem effectively.
In Table~\ref{utill}, we record the $U(\bm{x})$ and corresponding error at the last iteration. Still, PGD does not converge, and the error achieved by Exp-NUM is unsatisfactory. AGM overshoots the result while the restart methods perform well. AGM-fr achieves quite accurate $U(\bm{x})$ in both topologies.

\subsection{Performance with Different $d$}\label{exp3}

\begin{table}[tb]
  \centering
  \renewcommand\arraystretch{1}
  \caption{The achieved network utility value and error (in parentheses) when $U_s=\sum_{s\in S}x_s$ in G\'EANT and RF1221.\vspace{-1mm}}
   \begin{tabular}{|p{1.6cm}<{\centering}|p{2.4cm}<{\centering}|p{2.4cm}<{\centering}|}
    \hline
    \textbf{Algorithm} & \textbf{G\'EANT} & \textbf{RF1221}  \bigstrut\\
    \hline
    PGD & 657.09 (5890.11) & 1533.36 (21374.64)  \bigstrut\\
   \hline
    Exp-NUM & 6467.39 (79.81) & 22744.75 (163.26) \bigstrut\\
    \hline
    {AGM} & 6552.78 (-5.58)  & 22921.28 (-13.28)  \bigstrut\\
    \hline
     {AGM}-fr &  6547.199 (0.001) &  22907.97 (0.03) \bigstrut\\
    \hline
      {AGM}-gr &6547.19  (0.01) & 22906.87 (1.13) \bigstrut\\
      \hline
          LP Solver & 6547.20 & 22908.00  \bigstrut\\
    \hline
    \end{tabular}%
  \label{utill}%
    \vspace{-1mm}
\end{table}%

\begin{table}[tb]
  \centering
  \renewcommand\arraystretch{1}
  \caption{Average computation time per iteration with different number of flows $d$ when $U_s=\sum_{s\in S}\ln{(x_s+0.5)}$ in G\'EANT.\vspace{-1mm}}
   \begin{tabular}{|p{1.6cm}<{\centering}|p{1.6cm}<{\centering}|p{1.6cm}<{\centering}|p{1.6cm}<{\centering}|}
    \hline
    \textbf{Algorithm} & $\bm{d=}$\textbf{100} & \bm{$d=$}\textbf{1000} & \bm{$d=$}\textbf{10000} \bigstrut\\
    \hline
    PGD & 7.2$\times\text{10}^{\text{-5}}$s  & 1.7$\times\text{10}^{\text{-4}}$s & 1.1$\times\text{10}^{\text{-3}}$s \bigstrut\\
   \hline
    Exp-NUM & 9.3$\times\text{10}^{\text{-5}}$s  & 2.7$\times\text{10}^{\text{-4}}$s &1.3$\times\text{10}^{\text{-3}}$s \bigstrut\\
   \hline
   {AGM} & 8.0$\times\text{10}^{\text{-5}}$s & 2.0$\times\text{10}^{\text{-4}}$s & 1.2$\times\text{10}^{\text{-3}}$s\bigstrut\\
   \hline
     {AGM}-fr & 1.5$\times\text{10}^{\text{-4}}$s & 3.4$\times\text{10}^{\text{-4}}$s &1.8$\times\text{10}^{\text{-3}}$s\bigstrut\\
    \hline
     {AGM}-gr & 8.4$\times\text{10}^{\text{-5}}$s & 2.3$\times\text{10}^{\text{-4}}$s &  1.3$\times\text{10}^{\text{-3}}$s\bigstrut\\
    \hline
    \end{tabular}%
  \label{time}%
\end{table}%

We evaluate the impact of the number of flows $d$ on algorithm performance. Due to the space limitation, we take G\'EANT with $U_s=\sum_{s\in S}\ln{(x_s + 0.5)}$ for example, and run the algorithms when there are 100, 1000, 10000 flows in the network respectively. The results are similar for RF1221 and $U_s=\sum_{s\in S}x_s$.

Fig.~\ref{result2} shows the error at each iteration.
We've proven that the iteration complexity of AGM is $\Omega(\sqrt{d}/\varepsilon)$, so that AGM can keep good performance as $d$ increases. And the experiment results are consistent with the theoretical results. As show in  Fig.~\ref{result2}, different $d$ has no obvious effect on the convergence of AGM, it performs consistently well. Besides, although the iteration complexity of Exp-NUM is said to be nearly independent on $d$, our method AGM, AGM-fr and AGM-fr outperforms Exp-NUM with all $d$ settings.

In Table~\ref{time} we record the average computation time per iteration of different algorithms. As $d$ increases, the computation time per iteration also increases. With the same $d$, the time per iteration of PGD, Exp-NUM, and AGM are similar while Exp-NUM takes a little longer time for the computation of the exponential function, because they are all first-order methods. Thus, the difference in their convergence rates implies their different computation time to solve the problem. AGM-fr takes longer time because of the additional objective value computation when judging restart condition. Considering its fast and stable convergence, we think it's worth it.

\section{Limitations}\label{discussion}

\gai{Our method AGM aims at solving NUM problem with smooth utility functions. Although the most commonly studied $\alpha$-fair utility function is proven to be smooth, in general, however, utility functions are not necessarily smooth or even not concave, which prevents the use of our method.
Besides, one concern about the Softplus function is that the penalty value is although small, but still greater than zero when $\mathbf{A}_e\bm{x}\le c_e$. Such unwanted penalty can be mitigated by further moving the Softplus function towards the right of the horizontal axis or fine-tuning the parameter $\mu$.} \ggai{The convergence guarantee of AGM holds only when the number of links are much smaller than the number of flows in the network.}

\section{\vspace{-1mm}Conclusion}\label{conclusion}
\rev{In this paper, we propose a centralized, efficient and scalable accelerated gradient method (AGM) for the centralized NUM problem.} We discover the smoothness of \gai{the $\alpha$-fair} utility function, and formulate the NUM problem with smooth objective function, \rev{which not only benefits the existing method, but also enables the use of Nesterov's accelerated gradient method. We prove that AGM has the fastest} $O(d/t^2)$ convergence rate \rev{with respect to $t$}, and has $\Omega(\sqrt{d}/\sqrt{\varepsilon})$ iteration complexity \rev{which is scalable with respect to the number of flows $d$.} 
\rev{Experiment results are in good agreement with the theoretical results and demonstrate the great potential of AGM for resource allocation in large-scale centralized controlled networks.} \gai{In the future,  we will evaluate AGM with other smooth utility functions and penalty functions, and investigate its performance in real-world use cases.  \vspace{-2mm}}

\section*{Acknowledgment}
\gai{This work is supported in part by the National Key Research and Development Program of China under Grant No. 2018YFB1800400 and National Natural Science Foundation of China under Grant No. 62002009. We thank the anonymous reviewers for their valuable comments.}

\bibliographystyle{IEEEtran}
\bibliography{IEEEabrv,numref}

\begin{thebibliography}{10}
\providecommand{\url}[1]{#1}
\csname url@samestyle\endcsname
\providecommand{\newblock}{\relax}
\providecommand{\bibinfo}[2]{#2}
\providecommand{\BIBentrySTDinterwordspacing}{\spaceskip=0pt\relax}
\providecommand{\BIBentryALTinterwordstretchfactor}{4}
\providecommand{\BIBentryALTinterwordspacing}{\spaceskip=\fontdimen2\font plus
\BIBentryALTinterwordstretchfactor\fontdimen3\font minus
  \fontdimen4\font\relax}
\providecommand{\BIBforeignlanguage}[2]{{%
\expandafter\ifx\csname l@#1\endcsname\relax
\typeout{** WARNING: IEEEtran.bst: No hyphenation pattern has been}%
\typeout{** loaded for the language `#1'. Using the pattern for}%
\typeout{** the default language instead.}%
\else
\language=\csname l@#1\endcsname
\fi
#2}}
\providecommand{\BIBdecl}{\relax}
\BIBdecl

\bibitem{kelly1997}
F.~Kelly, ``Charging and rate control for elastic traffic,'' \emph{European
  transactions on Telecommunications}, vol.~8, no.~1, pp. 33--37, 1997.

\bibitem{10.1117/12.325891}
J.~Mo and J.~Walrand, ``{Fair end-to-end window-based congestion control},'' in
  \emph{Performance and Control of Network Systems II}, vol. 3530.\hskip 1em
  plus 0.5em minus 0.4em\relax SPIE, 1998, pp. 55 -- 63.

\bibitem{879343}
------, ``Fair end-to-end window-based congestion control,'' \emph{IEEE/ACM
  Transactions on Networking}, vol.~8, no.~5, pp. 556--567, 2000.

\bibitem{1665005}
A.~Eryilmaz and R.~Srikant, ``Joint congestion control, routing, and mac for
  stability and fairness in wireless networks,'' \emph{IEEE Journal on Selected
  Areas in Communications}, vol.~24, no.~8, pp. 1514--1524, 2006.

\bibitem{9139398}
N.~Karako\c{c}, A.~Scaglione, A.~Nedi\'{c}, and M.~Reisslein, ``Multi-layer
  decomposition of network utility maximization problems,'' \emph{IEEE/ACM
  Transactions on Networking}, vol.~28, no.~5, pp. 2077--2091, 2020.

\bibitem{8673789}
L.~Gu, D.~Zeng, S.~Tao, S.~Guo, H.~Jin, A.~Y. Zomaya, and W.~Zhuang,
  ``Fairness-aware dynamic rate control and flow scheduling for network utility
  maximization in network service chain,'' \emph{IEEE Journal on Selected Areas
  in Communications}, vol.~37, no.~5, pp. 1059--1071, 2019.

\bibitem{kelly1998}
F.~P. Kelly, A.~K. Maulloo, and D.~K.~H. Tan, ``Rate control for communication
  networks: shadow prices, proportional fairness and stability,'' \emph{Journal
  of the Operational Research Society}, vol.~49, no.~3, pp. 237--252, 1998.

\bibitem{811451}
S.~Low and D.~Lapsley, ``Optimization flow control. {I}. basic algorithm and
  convergence,'' \emph{IEEE/ACM Transactions on Networking}, vol.~7, no.~6, pp.
  861--874, 1999.

\bibitem{993307}
R.~La and V.~Anantharam, ``Utility-based rate control in the internet for
  elastic traffic,'' \emph{IEEE/ACM Transactions on Networking}, vol.~10,
  no.~2, pp. 272--286, 2002.

\bibitem{1664999}
D.~Palomar and M.~Chiang, ``A tutorial on decomposition methods for network
  utility maximization,'' \emph{IEEE Journal on Selected Areas in
  Communications}, vol.~24, no.~8, pp. 1439--1451, 2006.

\bibitem{10.1561/1300000007}
S.~Shakkottai and R.~Srikant, ``Network optimization and control,''
  \emph{Found. Trends Netw.}, vol.~2, no.~3, pp. 271--379, 2007.

\bibitem{16M1059011}
H.~Yu and M.~J. Neely, ``A simple parallel algorithm with an ${O}(1/t)$
  convergence rate for general convex programs,'' \emph{SIAM Journal on
  Optimization}, vol.~27, no.~2, pp. 759--783, 2017.

\bibitem{6756941}
A.~Beck, A.~Nedi\'{c}, A.~Ozdaglar, and M.~Teboulle, ``An ${O}(1/k)$ gradient
  method for network resource allocation problems,'' \emph{IEEE Transactions on
  Control of Network Systems}, vol.~1, no.~1, pp. 64--73, 2014.

\bibitem{Foundation2012Software}
O.~N. Foundation, ``Software-defined networking: The new norm for networks,''
  \emph{ONF White Paper}, 2012.

\bibitem{Jain2013B4}
S.~Jain, A.~Kumar, S.~Mandal, J.~Ong, L.~Poutievski, A.~Singh, S.~Venkata,
  J.~Wanderer, J.~Zhou, M.~Zhu \emph{et~al.}, ``B4: Experience with a
  globally-deployed software defined wan,'' in \emph{ACM SIGCOMM Computer
  Communication Review}, vol.~43, no.~4.\hskip 1em plus 0.5em minus 0.4em\relax
  ACM, 2013, pp. 3--14.

\bibitem{Hong2013Achieving}
C.-Y. Hong, S.~Kandula, R.~Mahajan, M.~Zhang, V.~Gill, M.~Nanduri, and
  R.~Wattenhofer, ``Achieving high utilization with software-driven wan,'' in
  \emph{ACM SIGCOMM Computer Communication Review}, vol.~43, no.~4.\hskip 1em
  plus 0.5em minus 0.4em\relax ACM, 2013, pp. 15--26.

\bibitem{7492926}
R.~Gupta, L.~Vandenberghe, and M.~Gerla, ``Centralized network utility
  maximization over aggregate flows,'' in \emph{2016 14th International
  Symposium on Modeling and Optimization in Mobile, Ad Hoc, and Wireless
  Networks (WiOpt)}, 2016, pp. 1--8.

\bibitem{8737600}
L.~Vigneri, G.~Paschos, and P.~Mertikopoulos, ``Large-scale network utility
  maximization: Countering exponential growth with exponentiated gradients,''
  in \emph{IEEE INFOCOM 2019 - IEEE Conference on Computer Communications},
  2019, pp. 1630--1638.

\bibitem{nesterov1983method}
Y.~E. Nesterov, ``A method for solving the convex programming problem with
  convergence rate ${O}(1/k^{2})$,'' in \emph{Dokl. akad. nauk Sssr}, vol. 269,
  1983, pp. 543--547.

\bibitem{NESTEROV2018907}
Y.~Nesterov and V.~Shikhman, ``Dual subgradient method with averaging for
  optimal resource allocation,'' \emph{European Journal of Operational
  Research}, vol. 270, no.~3, pp. 907--916, 2018.

\bibitem{MAL-050}
S.~Bubeck, ``Convex optimization: Algorithms and complexity,''
  \emph{Foundations and Trends\textsuperscript{\textregistered} in Machine
  Learning}, vol.~8, no. 3-4, pp. 231--357, 2015.

\bibitem{8064335}
Z.~Allybokus, K.~Avrachenkov, J.~Leguay, and L.~Maggi, ``Real-time fair
  resource allocation in distributed software defined networks,'' in \emph{2017
  29th International Teletraffic Congress (ITC 29)}, vol.~1, 2017, pp. 19--27.

\bibitem{5718026}
E.~Wei, A.~Ozdaglar, and A.~Jadbabaie, ``A distributed newton method for
  network utility maximization,'' in \emph{49th IEEE Conference on Decision and
  Control (CDC)}, 2010, pp. 1816--1821.

\bibitem{6831412}
M.~Zargham, A.~Ribeiro, and A.~Jadbabaie, ``Accelerated backpressure
  algorithm,'' in \emph{2013 IEEE Global Communications Conference (GLOBECOM)},
  2013, pp. 2269--2275.

\bibitem{6676846}
M.~Zargham, A.~Ribeiro, A.~Ozdaglar, and A.~Jadbabaie, ``Accelerated dual
  descent for network flow optimization,'' \emph{IEEE Transactions on Automatic
  Control}, vol.~59, no.~4, pp. 905--920, 2014.

\bibitem{nemirovskij1983problem}
A.~S. Nemirovskij and D.~B. Yudin, \emph{Problem complexity and method
  efficiency in optimization}.\hskip 1em plus 0.5em minus 0.4em\relax
  Wiley-Interscience, 1983.

\bibitem{6027859}
S.~Knight, H.~Nguyen, N.~Falkner, R.~Bowden, and M.~Roughan, ``The internet
  topology zoo,'' \emph{Selected Areas in Communications, IEEE Journal on},
  vol.~29, no.~9, pp. 1765 --1775, october 2011.

\bibitem{7417124}
C.~Filsfils, N.~K. Nainar, C.~Pignataro, J.~C. Cardona, and P.~Francois, ``The
  segment routing architecture,'' in \emph{2015 IEEE Global Communications
  Conference (GLOBECOM)}, 2015, pp. 1--6.

\bibitem{752159}
L.~Massoulie and J.~Roberts, ``Bandwidth sharing: objectives and algorithms,''
  in \emph{IEEE INFOCOM '99. Conference on Computer Communications.
  Proceedings. Eighteenth Annual Joint Conference of the IEEE Computer and
  Communications Societies. The Future is Now (Cat. No.99CH36320)}, vol.~3,
  1999, pp. 1395--1403 vol.3.

\bibitem{10.2307/2627851}
W.~I. Zangwill, ``Non-linear programming via penalty functions,''
  \emph{Management Science}, vol.~13, no.~5, pp. 344--358, 1967.

\bibitem{doi:10.1137/0804027}
M.~{\c{C}}. Pinar and S.~A. Zenios, ``On smoothing exact penalty functions for
  convex constrained optimization,'' \emph{SIAM Journal on Optimization},
  vol.~4, no.~3, pp. 486--511, 1994.

\bibitem{xu2013second}
X.~Xu, Z.~Meng, J.~Sun, L.~Huang, and R.~Shen, ``A second-order smooth penalty
  function algorithm for constrained optimization problems,''
  \emph{Computational optimization and applications}, vol.~55, no.~1, pp.
  155--172, 2013.

\bibitem{ZENIOS1995220}
\BIBentryALTinterwordspacing
S.~A. Zenios, M.~{\c{C}}. Pinar, and R.~S. Dembo, ``A smooth penalty function
  algorithm for network-structured problems,'' \emph{European Journal of
  Operational Research}, vol.~83, no.~1, pp. 220--236, 1995. [Online].
  Available:
  \url{https://www.sciencedirect.com/science/article/pii/037722179590601A}
\BIBentrySTDinterwordspacing

\bibitem{pmlr-v28-sutskever13}
I.~Sutskever, J.~Martens, G.~Dahl, and G.~Hinton, ``On the importance of
  initialization and momentum in deep learning,'' in \emph{International
  conference on machine learning}.\hskip 1em plus 0.5em minus 0.4em\relax PMLR,
  2013, pp. 1139--1147.

\bibitem{aaaaa}
B.~O'donoghue and E.~Candes, ``Adaptive restart for accelerated gradient
  schemes,'' \emph{Foundations of computational mathematics}, vol.~15, no.~3,
  pp. 715--732, 2015.

\bibitem{uhlig2006providing}
S.~Uhlig, B.~Quoitin, J.~Lepropre, and S.~Balon, ``Providing public intradomain
  traffic matrices to the research community,'' \emph{SIGCOMM Comput. Commun.
  Rev.}, vol.~36, no.~1, pp. 83--86, 2006.

\bibitem{10.1145/2829988.2787495}
R.~Hartert, S.~Vissicchio, P.~Schaus, O.~Bonaventure, C.~Filsfils, T.~Telkamp,
  and P.~Francois, ``A declarative and expressive approach to control
  forwarding paths in carrier-grade networks,'' \emph{SIGCOMM Comput. Commun.
  Rev.}, vol.~45, no.~4, pp. 15--28, 2015.

\end{thebibliography}

\end{document}